\documentclass[aps,twocolumn]{revtex4-2}
\usepackage{amsmath}
\usepackage{bm}
\usepackage{graphicx}
\usepackage{amssymb}
\usepackage{xcolor}

\begin{document}

\title{Collective Excitations and Optical Response of Ultrathin Carbon Nanotube Films}

\author{I. V. Bondarev}\email[Corresponding author: ]{ibondarev@nccu.edu}

\author{C.~M.~Adhikari}

\affiliation{Department of Mathematics \& Physics, North Carolina Central University, Durham, NC 27707, USA}

\begin{abstract}
We present a theoretical study of the collective quasiparticle excitations responsible for the electro\-magnetic response of ultrathin plane-parallel homogeneous periodic single-wall carbon nanotube arrays and weakly inhomogeneous single-wall carbon nanotube films. We show that in addition to varying film composition, the collective response can be controlled by varying the film thickness. For single-type nanotube arrays, the real part of the dielectric response shows a broad negative refraction band near a quantum interband transition of the constituent nanotube, whereby the system behaves as a hyper\-bolic metamaterial at higher frequencies than those classical plasma oscillations have to offer. By decreasing nanotube diameters it is possible to push this negative refraction into the visible region, and using weakly inhomogeneous multi-type nanotube films broadens its bandwidth.
\end{abstract}

\maketitle

\section{Introduction}\label{Sec:1}

Over the past years, the optical nanomaterials research has uncovered intriguing capabilities of carbon nanotubes (CNs).~These hollow graphene cylinders of one to a few nanometers in diameter and up to one centimeter in length~\cite{DressAvour01,Liu03,Doorn04} have landed themselves to attractive device applications~\cite{Baughman13,Ago99,AvourisNP08,Hertel10,Bond10,Bond14,Zettl06,Wim15,Htoon15,Krupke16,Mike17}. CNs have been successfully integrated into miniaturized electronic, electromechanical and chemical devices~\cite{Han10,Robel05,Vietmeyer07,Stranks11,Strano11}. A strong potential of CNs has been demonstrated in the fields of optical absorption and scattering spectroscopy~\cite{BoVi2006,BeEtAl2007,BlEtAl2013,RoEtAl2014,Bo15}, electron energy loss spectroscopy~\cite{MaEtal2003,Hage17}, quantum electron transport~\cite{Ago99,Mike17,GeBo2016}, and general collective electromagnetic (EM) response phenomena~\cite{An2005,DrSaJo2007,BoWoTa2009,Ando2010,Bo2011,BoAnt2012,Bo2012,Kono13,BoMe2014,Bo2014,Morimoto14,GAbacho15,Zaum16,MRSA2017,Kadochkin17,Kono18}. Carbon nanotube based reduced dimensionality materials such as periodic and quasiperiodic plane-parallel CN arrays and films, offer extraordinary stability, flexibility, and precise tunability of their physical properties by varying the density, diameter and chirality of constituent CNs. Thin and ultrathin periodically aligned arrays of single-wall CNs (SWCNs) and related systems are currently in the process of rapid experimental development~\cite{HerBond13,Kono16,FaEtAl2017,ChEtAl2017,HoEtAl2018,RoEtal2019,SC2020,JFan2020}.

\begin{figure}[b]
\includegraphics[width=1.0\linewidth]{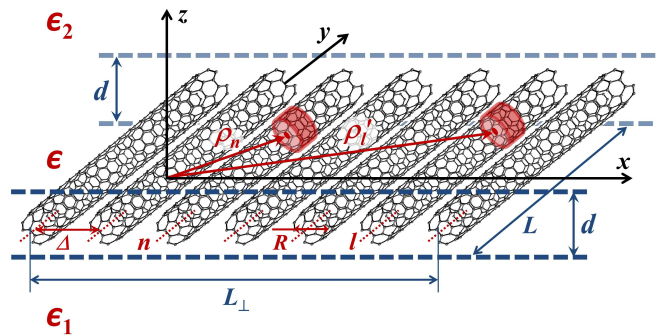}
\caption{Schematic of a plane-parallel, closely packed, periodically aligned array of SWCNs of radius $R$ and length $L$ embedded in a dielectric of thickness $d$. See text for details.}
\label{fig1}
\end{figure}

In this paper, we develop a theory for collective near-field interactions and associated EM response of ultrathin, closely packed, periodically aligned SWCN arrays (schematic shown in Fig.~\ref{fig1}). The key features that make this system interesting for theoretical study are the periodic CN alignment and the spatially periodic anisotropy associated with it. Additionally, the vertical confinement in dense ultrathin planar systems of finite thickness leads to the effective dimensionality reduction from 3D to 2D while still retaining the thickness as a parameter to represent the vertical size~\cite{BoMoSh2020,Bo2019,BoMoSh2018,BoSh2017}. This is the transdimensional (TD) regime~\cite{BolSh2019} --- neither 3D nor 2D but something in-between --- turning into 2D as thickness tends to zero, challenging to study what the 3D-to-2D continuous transition has to offer for new material functionalities.

The spatial anisotropy, periodic in-plane transverse inhomogeneity and vertical quantum confinement make the ultrathin array near fields strong and anisotropically nonlocal, adding both extra challenges in developing the problem theoretically and extra flexibility in designing CN films with desired EM properties experimentally. The properties of the CN array can be controlled if one knows how to control its collective excitations such as excitons, plasmons, and their mutual interactions. For example, plasmon generated near fields are known to affect spontaneous emission~\cite{BoLa2004,GAbacho15}, low-energy absorption~\cite{BoVi2006} and scattering~\cite{Bo15} by atomic type species near the CN surface. CN periodicity allows for plasmonic bands formation~\cite{GarsiaVidal97,Pitarkel98,GarsiaVidal98}, whereby periodic CN arrays and films should behave as epsilon-near-zero plasmonic metamaterials (MMs) in the near field, while still remaining strong light absorbers and polarizers in the far field~\cite{PolarCNSpectra,RoEtal2019}. Indeed, recent ellipsometry measurements on finite-thickness films of horizontally aligned self-assembled SWCNs have exhibited their tunable negative dielectric response for a sufficiently wide range of the photon excitation energies~\cite{RoEtal2019,SC2020}, thus paving the way for the highly anisotropic hyperbolic MM film (metasurface) development. Hyperbolic metamaterials (hMMs) and metasurfaces (MSs) can be used in a variety of applications~\cite{SmSc2003,CaiShalaev10,ZhelKiv12,KilBolShal13} such as optical sensing~\cite{SrEtal2016}, absorption~\cite{LaEtal2008}, cloaking~\cite{PeEtal2006}, and super-resolution imaging~\cite{LiEtal2007}. Being very stable and highly sensitive to the thickness, CN density, diameter and chirality variation, the finite-thickness parallel aligned SWCN arrays and films can additionally serve as a flexible multifunctional nanomaterial platform for single-molecule detection and manipulation~\cite{BoGu15,BoGeMe14,BoGeMe13}, including the near-field control of photoluminescence rate and directionality~\cite{BoLa2004,GAbacho15}, chemical reactivity~\cite{GaVidal15,GaVidal16}, and Casimir-Polder interactions~\cite{BoLa2005,BoLa2006}.

In this paper, we formulate the general Hamiltonian for the collective quasiparticle excitations in the ultrathin plane-parallel periodic SWCN array in the TD regime. This Hamiltonian is used to derive the in-plane dynamical dielectric response tensor of the system in the broad-band low-energy spectral range of microwave to visible ($\lesssim\!1\!-\!2$~eV). This is the domain where the intrinsic excitations, excitons and plasmons, are present in individual constituent (metallic and semiconducting) CNs ~\cite{Bo15,DrSaJo2007}. As opposed to the earlier 3D effective-medium response theories (known to be of limited validity at short CN separations) with only one single-tube excitation taken into account (the $6$~eV bulk graphite plasmon)~\cite{GarsiaVidal97,Pitarkel98,GarsiaVidal98}, our approach starts with the single-tube dynamical surface conductivity calculation as prescribed by the $\textbf{k}\!\cdot\!\textbf{p}$ band-structure simulation method for SWCNs within the Kubo linear-response theory~\cite{An2005}. To obtain the dielectric tensor of the TD array of SWCNs we use the low-energy plasmon response calculation technique proposed by one of us for finite-thickness metallic films with periodic cylindrical anisotropy~\cite{Bo2019}, combined with the many-particle Green's function formalism in the Matsubara formulation~\cite{Mahan2000}. With the single-tube conductivity in hand, we evaluate the collective polarization and obtain the EM response tensor for the periodic TD array of SWCNs. We also study the inhomogeneity effects for TD films composed of SWCN arrays with varied structural parameters, using the Maxwell-Garnett (MG) method~\cite{Mar2016} to obtain the EM response for weakly inhomogeneous quasiperiodic SWCN films reported experimentally~\cite{RoEtal2019,SC2020}.

The paper is organized as follows. Section~\ref{Sec:2} presents the model and formulates the Hamiltonian for the collective quasiparticle excitations in the TD plane-parallel periodic SWCN array. Section~\ref{Sec:3} uses this Hamiltonian to obtain the array collective polarization in terms of the surface conductivity of the individual constituent SWCN, followed by the in-plane EM response tensor derivation for the TD array, including the case of weakly inhomogeneous quasiperiodic SWCN films. Section~\ref{Sec:4} discusses our findings and numerical results. Conclusions are drawn in Sec.~\ref{Sec:5}. Mathematical details are collected in the two Appendices. We use the Gaussian units throughout unless otherwise stated.

\section{The Model of Collective Excitations}\label{Sec:2}

The model system we consider is presented in Fig.~\ref{fig1}. A~horizontal array of parallel aligned ($y$-direction) identical SWCNs of radius $R$ and length $L$ has the translational unit $\Delta$ and width $L_\perp$ ($x$-direction). The array is embedded in a dielectric of thickness $d$ to form a composite layer with the effective permittivity $\epsilon$, which is sandwiched between the substrate and superstrate with the dielectric permittivities $\epsilon_1$ and $\epsilon_2$, respectively. The structural parameters of the array obey the set of constraints as follows $2R\!\le\!\Delta\!\le\!d\!\ll\!L\!\sim\!L_\perp$ with $R$ and $\Delta$ being much less than the wavelength of the light radiation. The SWCN of the $(m,n)$ type ($n\!\le\!m$) has the radius $R\!=(\!\sqrt{3}\,b/2\pi)\sqrt{m^2+mn+n^2}$, where $b\!=\!1.42$~\AA\space is the C$\,$-C interatomic distance, with the electron charge density constrained by cylindrical symmetry to be uniformly distributed over its surface. The positive background of nuclei keeps the entire system electrically neutral.

We consider the most interesting case of $\epsilon\!\gg\epsilon_1+\epsilon_2$. In this case, with $d$ decreasing and getting less than the in-plane distance between a pair of electrons in the composite layer, due to the strong vertical confinement their Coulomb interaction is mostly contributed by the region outside of the layer, whereby the Coulomb interaction potential loses its $z$-coordinate dependence and turns into the Keldysh-Rytova (KR) potential~\cite{Ke1979}. This brings our system, the array, in the TD regime where the dimensionality of the system is reduced from 3D to 2D and the only remnant of the $z$-direction is the layer thickness $d$ to represent the size of the vertical confinement. In our case here, it is natural to take advantage of the homogeneous electron charge distribution over the SWCN cylindrical surfaces, whereby the pair electrostatic potential can be expressed in terms of that of two uniformly charged rings of radius $R$ each~\cite{Louie09}. For two such rings of the unit cells at points $\bm{\rho}_n$ and $\bm{\rho}_l^{\prime}$ of tubules $n$ and $l$ as shown in Fig.~\ref{fig1}, the electrostatic interaction potential can be written in terms of the symmetry adapted KR potential as follows~\cite{Bo2019}
\begin{equation}
V(\bm{\rho}_{n}\!-\!\bm{\rho}_l^\prime)=\frac{m^{\ast}}{N_{\rm 2D}LL_{\perp}}{\sum_\textbf{k}}^\prime\Big[\frac{\omega_p(k)}{k}\Big]^2
e^{i\textbf{k}\cdot(\bm{\rho}_{n}\!-\bm{\rho}_l^{\prime})}.
\label{V:eq1:IVB2019}
\end{equation}
Here, $\mathbf{k}\!=\!\mathbf{q}+\mathbf{k}_\perp$ is the in-plane electron quasimomentum with $\mathbf{q}$ and $\mathbf{k}_\perp$ representing its components in the $y$- and $x$-direction --- parallel and perpendicular to the CN alignment direction, respectively; $k\!=|\mathbf{k}|\!=\!\sqrt{q^2+k_\perp^2}$ where $q\!=\!2\pi n_{y}/L$, $n_{y}\!\!=0,\pm1,\pm2,...,\pm N/2$, $N\!\!=\!L/a$, $a$~is the CN lattice translation period, and $k_\perp\!\!=\!2\pi n_{x}/L_\perp$ where $n_{x}\!\!=\!0,\pm1,\pm2,...,\pm N_\perp/2$, $N_\perp$ is the total number of tubules in the parallel array. The summation is primed to eliminate the $\mathbf{k}\!=0$ term associated with the all-together electron displacement, $\bm{\rho}_n\!\ne\!\bm{\rho}_l^{\,\prime}$ for the $n\!=\!l$ case to exclude the self-interaction, and the orthogonality condition
\begin{eqnarray}
\delta_{\bm{\rho}_n\bm{\rho}_l^\prime}=\delta_{x_nx_l^\prime}\delta_{y_ny_l^\prime}=\frac{1}{N_\perp N}\sum_{\textbf{k}}e^{i\textbf{k}\cdot(\bm{\rho}_{n}-\bm{\rho}_l^\prime)}\label{normaliz1}\\
=\frac{1}{N_\perp}\hskip-0.6cm\sum_{~~~~~~k_{_{\!\perp}}=-\pi/\Delta}^{\pi/\Delta}\hskip-0.7cm e^{ik_{_{\!\perp}}\!(x_n-x_l^\prime)}\;
\frac{1}{N}\hskip-0.5cm\sum_{~~~~~q=-\pi/a}^{\pi/a}\hskip-0.5cm e^{iq(y_n-y_l^\prime)}\nonumber
\end{eqnarray}
is used for the Fourier expansion basis function set in the first Brillouin zone of the array, with its reciprocal given by
\begin{eqnarray}
\delta_{\textbf{k}\textbf{k}^\prime}=\delta_{k_{_{\!\!\perp}}{k_{_{\!\!\perp}}}^{\!\!\!\prime}}\,\delta_{qq^\prime}
=\frac{1}{N_\perp N}\sum_{\bm{\rho}_{n}}e^{i(\textbf{k}-\textbf{k}^\prime)\cdot\bm{\rho}_{n}}\hskip0.1cm\label{normaliz2}\\
=\frac{1}{N_\perp}\hskip-0.6cm\sum_{~~~~~x_n=-L_{_{\!\!\perp}}/2}^{L_{_{\!\!\perp}}/2}\hskip-0.7cm e^{i(k_{_{\!\perp}}\!-\,{k_{_{\!\!\perp}}}^{\!\!\!\prime}\,)x_n}\;
\frac{1}{N}\hskip-0.6cm\sum_{~~~~~y_n=-L\!/2}^{L\!/2}\hskip-0.6cm e^{i(q-q^\prime)y_n}.\nonumber
\end{eqnarray}
The quantity
\begin{equation}
\omega_p(k)=\sqrt{\frac{4\pi e^2N_{\rm 2D}}{m^{\ast}\epsilon\,d}\frac{2qRI_0(qR)K_0(qR)}{1+(\epsilon_1+\epsilon_2)/(k\epsilon\,d)}\,}
\label{plasmaFy}
\end{equation}
stands for the intraband plasma oscillation frequency for a general finite-thickness, cylindrically anisotropic, periodically aligned (metallic or semiconducting) array~\cite{Bo2019}. Here, $m^\ast$ is the electron effective mass, $N_{\rm 2D}$~$(=\!N_{\rm 3D}d)$ is the surface electron density, $I_0$ and $K_0$ are the zeroth-order modified cylindrical Bessel functions responsible for the correct normalization of the electron density distribution over cylindrical surfaces, to give for $R\!\rightarrow\!\infty$ (whereby $qRI_0(qR)K_0(qR)\!\rightarrow\!1/2$) the isotropic TD film plasma frequency studied previously~\cite{BoMoSh2020}.

\subsection{The Hamiltonian}

For an individual SWCN, absorption of a photon by a valence-band electron excites it to the conduction band to create an exciton. This makes a longitudinal polarization effect with an induced dipole moment directed predominantly along the nanotube axis due to a strongly suppressed SWCN polarizability in the perpendicular direction~\cite{Kozinsky,Louie95}, also known as the transverse depolarization~\cite{An2005,Tasaki98}. For a densely packed periodically aligned array of SWCNs, the induced inter-tubule dipole-dipole coupling should then result in the anisotropic collective polarization of the array (predominantly along the CN alignment direction), leading to the anisotropic dielectric response --- the focus of our study here.

The second quantized Hamiltonian of the free exciton (see, e.g., Refs.~\cite{Mahan2000,Haken}) on the SWCN surface can be written in the form
\begin{equation}
\hat{H}_{ex}\!=\!\sum_{s,q}E_s(q)B^\dag_{s,q}B_{s,q},
\label{Hex}
\end{equation}
where the operators $B^\dag_{s,q}$ and $B_{s,q}$ create and annihilate, respectively, the exciton with the longitudinal quasimomentum $q$ ($y$-direction in Fig.~\ref{fig1}) in the $s$-subband, which for the $(m,n)$-type SWCN is associated with the quantized circumferential momentum component $\hbar s/R$, $s\!=\!1,2,...,m\,(\ge\!n)$~\cite{Bo2012,Bo2014}. These operators satisfy the standard bosonic commutation relations
\[
\big[B_{s,q},B^\dag_{s^\prime\!\!,q^\prime}\big]=\delta_{ss^\prime}\delta_{qq^\prime},~~\big[B_{s,q},B_{s^\prime\!\!,q^\prime}\big]=0
\]
and can be converted into their coordinate counterparts using Eq.~(\ref{normaliz1}), to obtain
\begin{equation}
B^\dag_{s,y_n}\!\!=\frac{1}{\sqrt{N}}\sum_{q}B^\dag_{s,q}e^{iqy_n},~~B_{s,y_n}\!\!=\big(B^\dag_{s,y_n}\big)^\dag
\label{Bkf}
\end{equation}
for the $n$-th SWCN of the array. The exciton energy is
\begin{equation}
E_s(q)=E_{exc}^{(f)}(s)+\frac{\hbar^2q^2}{2M_{ex}(s)}\,,
\label{Ef}
\end{equation}
where the first term is the excitation energy of the $f$-internal-state exciton, given by $E_{exc}^{(f)}(s)\!=\!E_g(s)+E_b^{(f)}(s)$ with $E_b^{(f)}(s)$ being its (negative) binding energy and $E_g$ representing the band gap, and the second term is the translational kinetic energy of the exciton with the total effective mass $M_{ex}$ (the sum of the subband-dependent electron and hole effective masses). The exciton on the $n$-th SWCN of the array is produced by the absorption of a photon of the external EM radiation. This initiates the interband dipole electronic transitions $\langle s,f|\hat{\textbf{d}}(\bm{\rho}_n)|0\rangle$ to create the dipole polarization, an observable quantity proportional to $|\!\sum_{f,\bm{\rho}_n}\!\langle s,f|\hat{\textbf{d}}(\bm{\rho}_n)|0\rangle|^2$, mainly along the CN axis. Only the ground-internal-state (longest-lived) excitons are considered in what follows and so the $f$-index will be omitted for brevity.

The total Hamiltonian of our model system in Fig.~\ref{fig1} can now be written as
\begin{equation}
\hat{H}=\hat{H}_{ex}\!+\hat{H}_{int}\,,
\label{Eq:H}
\end{equation}
where the second term is responsible for the collective excitations of the SWCN array originally created by the external EM radiation on an individual constituent SWCN. To obtain this part, we perform the series expansion of the electron interaction potential~(\ref{V:eq1:IVB2019}) about the equilibrium positions of the electron density distribution. Intro\-ducing $\bm{\rho}_n\!+\bm{\beta}_n$ for the electron ring position displaced by a vector $\bm{\beta}_n\!=\!\bm{\beta}(\bm{\rho}_n)$ from the point $\bm{\rho}_n$ of the $n$-th SWCN electron density distribution (and same for ${\bm{\rho}}_l^\prime$; see Fig.~\ref{fig1}), the exponential factor in Eq.~(\ref{V:eq1:IVB2019}) is expanded to the third nonvanishing term assuming that $|{\bm{\rho}}_n\!-{\bm{\rho}}_l^\prime|\!>\!|\bm{\beta}_n\!-\bm{\beta}_l^\prime|$, to give
\[
e^{i\textbf{k}\cdot\bm{\rho}_{nl}}\big[1 + i\textbf{k}\cdot\bm{\beta}_{nl}-\frac{1}{2}\big(\textbf{k}\cdot\bm{\beta}_{nl}\big)^2\big].
\]
Here, $\bm{\rho}_{nl}\!=\!\bm{\rho}_n-\bm{\rho}^\prime_l$ and $\bm{\beta}_{nl}\!=\bm{\beta}_n-\bm{\beta}^\prime_l$ with the primes still to indicate that the self-interaction in the $n\!=l$ case is to be excluded. For achiral $(m,0)$ and $(m,m)$ type SWCNs (zigzag and armchair type, respectively) we choose to focus on here, the second term in square brackets vanishes due to the presence of the inversion symmetry. This is generally not true for chiral SWCNs, where it is nonzero and turns out to be the most important one. This case will be addressed elsewhere separately.

The extra electron potential energy that comes out of Eq.~\eqref{V:eq1:IVB2019} due to the relative electron ring displacement~$\bm{\beta}_{nl}$, is given by $\sum_{\bm{\rho}_{n\!},\bm{\rho}_{l}^\prime}\!\!\big[V(\bm{\rho}_{nl}+\bm{\beta}_{nl})-V(\bm{\rho}_{nl})\big]$ where the summation runs over all tubules of the array and over their individual unit cells with the self-interaction excluded. With the series expansion above and the self-interaction terms dropped, this becomes
\[
\frac{m^{\ast}}{N_{\rm 2D}LL_{\perp}}\!\!\!\!\!{\sum_{~~~\textbf{k},\bm{\rho}_{n\!},\bm{\rho}_{l}^\prime}}\!\!\!\!\!e^{i\textbf{k}\cdot\bm{\rho}_{nl}}
\Big[\frac{\omega_p(k)}{k}\Big]^2\big(\textbf{k}\cdot\bm{\beta}_{n}\big)\big(\textbf{k}\cdot\bm{\beta}_{l}^\prime\big),
\]
and can further be rewritten as follows
\begin{eqnarray}
\frac{1}{v_0}\hskip-0.6cm\sum_{~~~~~\textbf{k},\bm{\rho}_{n\!},\bm{\rho}_{l}^\prime,\mu,\nu}
\hskip-0.7cm e^{i\textbf{k}\cdot\bm{\rho}_{n}}\sqrt{\frac{m^{\ast\,}d}{N_{\rm 2D}N_{\perp}N}}\;\omega_p(k)
\Big[\beta_{n\mu}T_{\mu\nu}(\textbf{k})\beta_{l\nu}^\prime\hskip0.5cm\label{extraPotEn}\\
+\,\frac{1}{2}\,\beta_{n\mu}\beta_{l\nu}^\prime\delta_{\mu\nu}\Big]\sqrt{\frac{m^{\ast\,}d}{N_{\rm 2D}N_{\perp}N}}\;
\omega_p(k)\,e^{-i\textbf{k}\cdot\bm{\rho}_{l}^\prime}.\hskip0.9cm\nonumber
\end{eqnarray}
Here, $v_0\!=a\Delta d$ is the elementary cell volume of the array and the second-rank tensor
\begin{equation}
T_{\mu\nu}(\textbf{k})=\frac{k_\mu k_\nu}{k^2}-\frac{\delta_{\mu\nu}}{2}\,,~~~\mu,\nu=x,y
\label{Tmunu}
\end{equation}
represents the Fourier transform of the pair dipole-dipole interaction between the SWCNs in the array~\cite{Mahan2000}. The second term in the square brackets equals zero. This can be proven by first summing it up over $\bm{\rho}_l^\prime$ using Eq.~(\ref{normaliz2}), followed by summation over $\textbf{k}$, which leads to $\omega_p^2(0)\!=\!0$.

The dipole-dipole interaction energy in Eq.~(\ref{extraPotEn}) can be used to obtain the $\hat{H}_{int}$ part of the total Hamiltonian~(\ref{Eq:H}). Consider the real vector quantity
\begin{equation}
\textbf{d}(\bm{\rho}_{n})\!=\!\sqrt{\!\frac{m^{\ast\,}d}{4\pi N_{\rm 2D}}}\,\omega_p(k)\bm{\beta}(\bm{\rho}_{n})\!
=\!\sqrt{\!\frac{m^{\ast\,}d}{4\pi N_{\rm 2D}N}}\,\omega_p(k)\beta\textbf{e}(\bm{\rho}_{n})
\label{drhonEigen}
\end{equation}
with the unit vector $\textbf{e}(\bm{\rho}_{n})\!=\!\bm{\beta}(\bm{\rho}_n)/\beta(\bm{\rho}_{n})$ to define the direction of the electron ring displacement $\bm{\beta}(\bm{\rho}_{n})$ and
\[
\beta=\sqrt{|\!\sum_{\bm{\rho}_{n}}\bm{\beta}(\bm{\rho}_{n})|^2}\approx\sqrt{\sum_{\bm{\rho}_{n}}|\bm{\beta}(\bm{\rho}_{n})|^2}=\sqrt{N}\beta(\bm{\rho}_{n})
\]
to represent the net electron density displacement (consistent with the random phase approximation and equivalency of rings) for the entire $n$-th SWCN. Let Eq.~(\ref{drhonEigen}) be a matrix element $\langle s,\bm{\rho}_{n}|\hat{\textbf{d}}(\bm{\rho}_{n})|0\rangle$ of the (Hermitian) induced dipole moment operator defined in the second quantization form as
\begin{eqnarray}
\hat{\textbf{d}}(\bm{\rho}_{n})\!=\!\sqrt{\!\frac{m^{\ast\,}d}{4\pi N_{\rm 2D}N}}\,\omega_p(k)\!
\sum_s\langle s|\hat{\bm{\beta}}|0\rangle\big(B_{s,\bm{\rho}_{n}}\!\!+\!B^\dag_{s,\bm{\rho}_{n}}\big),\hskip0.5cm\label{drhonOper}\\
\langle s|\hat{\bm{\beta}}|0\rangle^\dag\!=\langle0|\hat{\bm{\beta}}|s\rangle^\ast\!=\langle0|\hat{\bm{\beta}}|s\rangle\!=\langle s|\hat{\bm{\beta}}|0\rangle,\hskip1.4cm\nonumber
\end{eqnarray}
where $\hat{\bm{\beta}}$ is the quantum displacement operator to represent the classical displacement $\beta\textbf{e}(\bm{\rho}_{n})$. Then the Fourier transform
\[
\hat{\textbf{d}}(\textbf{k})=\frac{1}{\sqrt{N_\perp N}}\sum_{\bm{\rho}_{n}}\hat{\textbf{d}}(\bm{\rho}_{n})e^{-i\textbf{k}\cdot\bm{\rho}_{n}},~~
\hat{\textbf{d}}^\dag(\textbf{k})\!=\hat{\textbf{d}}(-\textbf{k}),
\]
obtained using Eqs.~(\ref{normaliz1}) and (\ref{normaliz2}), takes the form
\begin{eqnarray}
\hat{\textbf{d}}(\textbf{k})\!=\sqrt{\!\frac{m^{\ast\,}d}{4\pi N_{\rm 2D}N}}\,\omega_p(k)\!\sum_s\langle s|\hat{\bm{\beta}}|0\rangle
\big(B_{s,\textbf{k}}\!+B^\dag_{s,-\textbf{k}}\big),\hskip0.5cm\label{dkOper}\\
B_{s,\textbf{k}}=\frac{1}{\sqrt{N_\perp N}}\sum_{\bm{\rho}_{n}}B_{s,\bm{\rho}_{n}}e^{-i\textbf{k}\cdot\bm{\rho}_{n}}\hskip1.5cm\nonumber
\end{eqnarray}
with the creation and annihilation operators of the same meaning as those in Eq.~(\ref{Hex}), and Eq.~(\ref{extraPotEn}) suggests that
\begin{equation}
\hat{H}_{int}=\frac{4\pi}{v_0}{\sum_{\textbf{k},\mu,\nu}}\hat{d}_\mu(\textbf{k})T_{\mu\nu}(\textbf{k})\hat{d}_\nu(-\textbf{k}).
\label{Eq:HintGen1}
\end{equation}
More explicitly
\begin{eqnarray}
\hat{H}_{int}=\frac{1}{2}\!\sum_{\textbf{k},s,s^\prime}\!\!V_{ss^\prime}(\textbf{k})\big(B_{s,\textbf{k}}\!+\!B^\dag_{s,-\textbf{k}}\big)
\big(B_{s^\prime\!,-\textbf{k}}\!+\!B^\dag_{s^\prime\!,\textbf{k}}\big),\hskip0.5cm\label{Eq:HintGen2}\\
V_{ss^\prime}(\textbf{k})=\frac{2m^{\ast}\omega_p^2(k)d}{N_{\rm 2D}Nv_0}\sum_{\mu,\nu}\langle 0|\hat{\beta}_\mu|s\rangle T_{\mu\nu}(\textbf{k})\langle s^\prime|\hat{\beta}_\nu|0\rangle.\hskip0.75cm\nonumber
\end{eqnarray}
Here, different spatial component matrix elements of the displacement operator $\hat{\bm{\beta}}$ in $V_{ss^\prime}$ are different, in general.

In the case where the (dominant) longitudinal induced SWCN polarization is only taken into consideration and the transverse one is neglected, one has $\beta_{n\mu}\!=\beta_{ny}\delta_{y\mu}$ and $\beta_{l\nu}^\prime\!=\beta_{ly}^\prime\delta_{y\nu}$ in Eq.~(\ref{extraPotEn}). This, after summing up over $x_n$ and $x_l^\prime$ with Eq.~(\ref{normaliz2}), projects it on the $y$-direction to give
\begin{eqnarray}
\frac{1}{v_0}\!\sum_{q,y_{n\!},y_{l}^\prime}\!e^{iq\cdot y_{n}}\sqrt{\frac{m^{\ast\,}d}{N_{\rm 2D}N}}\,\omega_p(q)\,\beta_{ny}\hskip2.5cm\nonumber\\[-0.25cm]
\hskip2.5cm\times\,T_{yy}(q)\,\beta_{ly}^\prime\sqrt{\frac{m^{\ast\,}d}{N_{\rm 2D}N}}\,\omega_p(q)\,e^{-iq\cdot y_{l}^\prime},\nonumber
\end{eqnarray}
where $T_{yy}(q)\!=T_{yy}(q,k_\perp\!\!=\!0)=1/2$ according to Eq.~(\ref{Tmunu}) and $\omega_p(q)\!=\omega_p(q,k_\perp\!\!=\!0)$, thus yielding
\begin{eqnarray}
\hat{H}_{int}=\frac{2\pi}{v_0}{\sum_{q}}\,\hat{d}_y(q)\hat{d}_y(-q)=\frac{2\pi}{v_0}{\sum_{q}}\,\hat{d}_y(q)\hat{d}_y^{\,\dag}(q),\hskip0.9cm\label{Eq:Hint1q}\\
\hat{d}_y(q)\!=\sqrt{\!\frac{m^{\ast}d}{4\pi N_{\rm 2D}N}}\,\omega_p(q)\!\sum_s\,\langle s|\hat{\beta}_y|0\rangle\big(B_{s,q}\!+B^\dag_{s,-q}\big)\hskip0.5cm\nonumber
\end{eqnarray}
for Eq.~(\ref{Eq:HintGen1}) and
\begin{eqnarray}
\hat{H}_{int}=\frac{1}{2}\sum_{s,q}V_{ss}(q)\big(B_{s,q}\!+B^\dag_{s,-q}\big)\big(B_{s,-q}\!+B^\dag_{s,q}\big),\hskip0.5cm\label{Eq:Hint2q}\\
V_{ss}(q)=\frac{m^{\ast}\omega_p^2(q)d}{N_{\rm 2D}Nv_0}\,|\langle s|\hat{\beta}_y|0\rangle|^2.\hskip1.75cm\nonumber
\end{eqnarray}
for Eq.~(\ref{Eq:HintGen2}), respectively. In this latter case, $V_{ss^\prime}$ takes the diagonal form $V_{ss}\delta_{ss^\prime}$ to only include the collinear-anticollinear dipole moment coupling between the nano\-tubes of the array. This is the case of practical importance we focus on in what follows.

\subsection{The Dispersion of Collective Excitations}

With the interaction term of Eq.~(\ref{Eq:Hint2q}), the total Hamiltonian (\ref{Eq:H}) of the SWCN array is of the form
\begin{eqnarray}
\hat{H}=\!\sum_{s,q}\Big[E_s(q)B^\dag_{s,q}B_{s,q}\hskip1.2cm\label{Eq:Htotfin}\\
+\,\frac{1}{2}\,V_{ss}(q)\big(B_{s,q}\!+B^\dag_{s,-q}\big)\big(B_{s,-q}\!+B^\dag_{s,q}\big)\Big],\nonumber
\end{eqnarray}
which can be diagonalized exactly by means of the canonical transformation technique~\cite{Bogoliub}. More specifically, the Hamiltonian such as this can be brought to the form
\begin{equation}
\hat{H}=\!\sum_{s,q}\hbar\omega_s(q)\xi_s^\dag(q)\xi_s(q)+E_0
\label{Eq:Htotdiag}
\end{equation}
by redefining the original exciton operators as follows
\begin{eqnarray}
B_{s,q}\!=u_s(q)\,\xi_s(q)+v_s(q)\,\xi^\dag_s(-q),\label{transform1}\\
|u_s(q)|^2-|v_s(q)|^2=1,\hskip1.0cm\label{ftransf}
\end{eqnarray}
in terms of the linear superposition of the collective boson excitation operators $\xi_s(q)$ representing the \emph{eigen} states of the entire system, the array. The reciprocal of this is
\begin{eqnarray}
\xi_s(q)=u^\ast_s(q)B_{s,q}-v_s(q)B^\dag_{s,-q}\label{xi}\\
\big[\xi_s(q),\xi^\dag_{s^\prime}(q^\prime)\big]=\delta_{ss^\prime}\delta_{qq^\prime}.\hskip0.5cm
\label{xicommute}
\end{eqnarray}
The unknown transformation coefficients $u_s(q)$ and $v_s(q)$, and the eigen state energies $\hbar\omega_s(q)$ can be found from the operator identity
\begin{equation}
\hbar\omega_s(q)\xi_s(q)=\big[\xi_s(q),\,\hat{H}\big],
\label{identity}
\end{equation}
followed by finding the vacuum (no excitations present) energy $E_0$ from the comparison of Eqs.~(\ref{Eq:Htotfin}) and (\ref{Eq:Htotdiag}) after the eigen state energies and the transformation coefficients have been obtained.

Putting Eqs.~(\ref{xi}) and (\ref{Eq:Htotfin}) into Eq.~(\ref{identity}) and equating the coefficients in front of the exciton creation and annihilation operators on the left and right side, one obtains the set of two simultaneous algebraic equations as follows
\begin{eqnarray}
\big[\hbar\omega_s(q)-E_s(q)-V_{ss}(q)\big]u_s(q)-V_{ss}(q)v_s(q)=0,\hskip0.5cm\nonumber\\[-0.15cm]
\label{equations}\\[-0.15cm]
V_{ss}(q)u_s(q)+\big[\hbar\omega_s(q)+E_s(q)+V_{ss}(q)\big]v_s(q)=0.\hskip0.55cm\nonumber
\end{eqnarray}
Being supplemented by the extra constraint (\ref{ftransf}), these simultaneous equations define the transformation coefficients $u_s(q)$ and $v_s(q)$ uniquely. The condition of the nontrivial solution existence requires that the determinant of Eq.~(\ref{equations}) be equal to zero. This gives the collective excitation eigen energies in the form
\begin{equation}
\hbar\omega_s(q)=\sqrt{E_s^2(q)+2E_s(q)V_{ss}(q)}
\label{eigenen}
\end{equation}
with $E_s(q)$ given by Eq.~(\ref{Ef}), $V_{ss}(q)$ defined in Eq.~(\ref{Eq:Hint2q}), and $s\!=\!1,2,...,m\,(\ge\!n)$ representing the excitation subbands for the periodic array composed of the $(m,n)$ type SWCNs. Using Eq.~(\ref{eigenen}), the transformation coefficients come out of Eq.~(\ref{equations}) in the compact form as follows
\begin{equation}
u_s(q)\!=\!\frac{E_s(q)\!+\!\hbar\omega_s(q)}{2\sqrt{E_s(q)\hbar\omega_s(q)}},~v_s(q)\!=\!\frac{E_s(q)\!-\!\hbar\omega_s(q)}{2\sqrt{E_s(q)\hbar\omega_s(q)}}\,.
\label{coefficients}
\end{equation}
Finally, substituting Eq.~(\ref{xi}) in Eq.~(\ref{Eq:Htotdiag}), using Eqs.~(\ref{eigenen}) and (\ref{coefficients}) for simplifications, and comparing the end result with Eq.~(\ref{Eq:Htotfin}), one obtains the vacuum state energy of the array in the form
\begin{equation}
E_0=\frac{1}{2}\sum_{s,q}\big[\hbar\omega_s(q)-E_s(q)\big].
\label{E0}
\end{equation}

Equations~(\ref{Eq:Htotdiag}), (\ref{xi}), (\ref{xicommute}), and (\ref{eigenen})-(\ref{E0}) solve the dispersion relation problem for the intrinsic collective excitations in the ultrathin periodic SWCN array. One can now see that the excited \emph{eigen} states of the SWCN array involve not only the single-tube excitons of energy $E_s(q)$ as one might expect naturally, but also the array collective plasma oscillations of energy $\hbar\omega_p(q)$ given by Eq.~(\ref{plasmaFy}) are involved, which in the TD regime can be effectively controlled by varying the array thickness~\cite{Bo2019}.

\section{The Array Dielectric Response}\label{Sec:3}

Knowing the dispersion of the excited collective quasiparticle \emph{eigen} states of the planar SWCN array, allows one to calculate its in-plane dielectric response tensor. A relatively easy way to do this is to use the complex-frequency finite-temperature Matsubara Green's function formalism (see, e.g., Refs.~\cite{Mahan2000,Abr1975}). All measurable collective quantities, such as conductivities and susceptibilities, are represented by correlation functions expressed for causality in terms of the retarded Green's functions, which can be obtained from their equivalent complex-frequency Matsubara counterparts with $i\omega\!=\!2n\pi/\hbar\beta$ for bosons and $i\omega\!=\!(2n+1)\pi/\hbar\beta$ for fermions ($n$ is an integer and $\beta=1/k_BT$) by simply changing $i\omega$ to $\omega+i\delta$ with $\delta$ representing an infinitesimal energy relaxation rate~\cite{Mahan2000}.

\subsection{The Collective Polarization}

We proceed with the polarizability tensor calculation for the SWCN array within the Matsubara formalism. This is given by the correlator of the SWCN induced dipole moment operator in Eq.~(\ref{Eq:Hint1q}) with itself, and this only includes one CN related component, $P_{yy}(q,i\omega)$, since the minor polarization perpendicular to the CN axis is agreed to be neglected. Specifically,
\begin{equation}
P_{yy}(q,i\omega)=-\frac{1}{\hbar}\int_{0}^{\hbar\beta}\!\!\!\!\!\!\!d\tau\,e^{i\omega\tau}\langle T_\tau\hat{d}_y(q,\tau)\hat{d}_y(-q,0)\rangle,
\label{Eq:Fij0}
\end{equation}
where the bracket $\langle\,\cdots\rangle$ notates the thermodynamical averaging, $\tau\,(=\!it)$ is the complex time, and $T_\tau$ is the order\-ing operator to place the (Heisenberg picture) dipole moment operators with the earliest $\tau$ to the right.

A straightforward way to calculate Eq.~(\ref{Eq:Fij0}) is by using Eq.~(\ref{transform1}) to write the single-tube induced dipole operators in terms of their collective excitation counterparts, followed by the thermodynamical averaging with the total Hamiltonian (\ref{Eq:Htotdiag}) in the collective excitation Hilbert space. However, for practical use purposes it is instructive to find the connection between the single-tube excitations and the collective excitations of the entire SWCN array. To this end, we first do the averaging in Eq.~(\ref{Eq:Fij0}) with the unperturbed Hamiltonian~(\ref{Hex}) to obtain the \emph{noninteracting} SWCN array polarizability (see Appendix~\ref{Appx1})
\begin{equation}
P_{yy}^{(0)}(q,i\omega)=\frac{m^\ast\omega^2_p(q)d}{4\pi N_{\rm 2D}N}\sum_s\frac{2E_s|\langle s|\hat{\beta}_y|0\rangle|^2}{(i\hbar\omega)^2-E_s^2}\,.
\label{Eq:Fyy0:Simp}
\end{equation}
Introducing the CN dynamical polarizability function
\begin{equation}
\alpha_{yy}(q,i\omega)=-\frac{e^2}{L}\sum_s\frac{2E_s|\langle s|\hat{\beta}_y|0\rangle|^2}{(i\hbar\omega)^2-E_s^2}
\label{alfayy}
\end{equation}
links Eq.~(\ref{Eq:Fyy0:Simp}) to the optical response of an isolated CN (longitudinal polarizability per unit length) to give
\begin{equation}
P_{yy}^{(0)}(q,i\omega)=-\frac{m^\ast\omega^2_p(q)v_0}{4\pi e^2N_{\rm 2D}\Delta}\,\alpha_{yy}(q,i\omega).
\label{P0yyalfa}
\end{equation}
Alternatively, this can be expressed in an equivalent form in terms of the isolated CN longitudinal conductivity~\cite{An2005}
\begin{equation}
\sigma_{yy}(q,i\omega)=\frac{\hbar e^2}{2\pi RL}\sum_s\frac{-2\hbar\omega|\langle s|\hat{v}_y|0\rangle|^2}{E_s\big[(i\hbar\omega)^2-E_s^2\big]}\,.
\label{Eq:sigma:Ando}
\end{equation}
Using the relations between the velocity and coordinate displacement matrix elements
\begin{equation}
\langle s|\hat{v}_y|0\rangle\!=\!\langle s|\frac{d\hat{\beta}_y}{d\tau}|0\rangle\!=
\!-\frac{1}{\hbar}\langle s|[\hat{\beta}_y,\hat{H}_{ex}]|0\rangle\!=\!\frac{E_s}{\hbar}\,\langle s|\hat{\beta}_y|0\rangle
\label{veldisp}
\end{equation}
and the fact that~\cite{BoLa2005,BoLa2006}
\begin{equation}
\alpha_{yy}(q,i\omega)=\frac{2\pi R}{\omega}\,\sigma_{yy}(q,i\omega),
\label{alfasigma}
\end{equation}
one then obtains
\begin{equation}
P_{yy}^{(0)}(q,i\omega)=-\frac{m^\ast\omega^2_p(q)v_0}{4\pi e^2N_{\rm 2D}\Delta}\frac{2\pi R}{\omega}\,\sigma_{yy}(q,i\omega).
\label{Eq:Pyy0v}
\end{equation}

Next, being performed with the total Hamiltonian (\ref{Eq:H}), the thermodynamical averaging in Eq.~(\ref{Eq:Fij0}) gives the collective polarizability of the dipole-dipole \emph{coupled} SWCN array. This can be evaluated by a diagrammatic expansion in which the first term, given by Eq.~(\ref{Hex}), is treated as unperturbed and the second, given by Eqs.~(\ref{Eq:HintGen1})-(\ref{Eq:Hint2q}), is the perturbation. For the former, the noninteracting SWCN array polarizability is already known and given by Eqs.~(\ref{P0yyalfa}) and (\ref{Eq:Pyy0v}). An important feature of the latter is its separability, as can be seen from Eqs.~(\ref{Eq:HintGen1}) and (\ref{Eq:Hint1q}), so that the perturbation factors into components summed up individually. The self-energy term for the perturbation occurs in the first order with the self-energy merely equal to $2\pi/v_0$, as per Eq.~(\ref{Eq:Hint1q}). The higher-order terms of the diagrammatic expansion just produce multiples of this term. The overall result takes the form of the Dyson equation as follows (see Appendix~\ref{Appx2})
\[
P_{yy}(q,i\omega)=P_{yy}^{(0)}(q,i\omega)+\frac{2\pi}{v_0}P_{yy}^{(0)}(q,i\omega)P_{yy}(q,i\omega),
\]
which can be easily solved to give
\begin{equation}
P_{yy}(q,i\omega)=\frac{P_{yy}^{(0)}(q,i\omega)}{1-2\pi P_{yy}^{(0)}(q,i\omega)/v_0}\,.
\label{Pyy}
\end{equation}

This is the collective array polarization associated with the \emph{interband} electronic transitions to create excitons on individual SWCNs embedded in a finite-thickness dielectric layer. There is also the low-energy polarization part contributed by the \emph{intraband} transitions of harmonically bound charges on the CN surface~\cite{GAbacho15,Bo2019}. The total collective polarization of the CN array is to include both intra- and interband term on equal footing since both of them are associated with an induced oscillating dipole moment --- in the lower- and higher-frequency spectral range, respectively, --- initially created by external EM radiation on an individual CN and spread out over the array thereafter. This can be formally done by redefining
\begin{eqnarray}
\sigma_{yy}(q,i\omega)\longrightarrow\sigma_{yy}^{intra}(q,i\omega)+\sigma_{yy}^{inter}(q,i\omega)
\label{redef}
\end{eqnarray}
in Eq.~(\ref{Eq:Pyy0v}), where
\begin{equation}
\sigma_{yy}^{intra}(q,i\omega)=\frac{e^2N_{\rm 2D}}{m^\ast}\frac{\omega}{\omega^2+(v_Fq)^2}
\label{sigmaintra}
\end{equation}
generally represents the single-tube intraband conductivity term with $v_F$ being the Fermi velocity~\cite{Ando2009}, and $\sigma_{yy}^{inter}$ is the interband term still given by Eq.~(\ref{Eq:sigma:Ando}). The SWCN polarizability in Eq.~(\ref{P0yyalfa}) has to be redefined per Eqs.~(\ref{alfasigma}) and (\ref{redef}) as well.

\subsection{The Dielectric Response}

Using Eq.~(\ref{Pyy}) with $P_{yy}^{(0)}$ of Eq.~(\ref{Eq:Pyy0v}) in the general EM response expression (Gaussian units)
\begin{equation}
\frac{\varepsilon_{yy}(q,i\omega)}{\epsilon}=1+\frac{4\pi}{v_0}\,P_{yy}(q,i\omega),
\label{epsgeneral}
\end{equation}
after simplifications one obtains the SWCN array dynamical dielectric response in the form
\[
\frac{\varepsilon_{yy}(q,i\omega)}{\epsilon}=1-\!\frac{2f_{\rm CN\,}\sigma_{yy}(q,i\omega)}{f_{\rm CN\,}\sigma_{yy}(q,i\omega)+\omega e^2N_{\rm 2D}R/m^\ast\omega^2_p(q)d}\,,
\]
where $f_{\rm CN}\!=\!N_\perp V_{\rm CN}/V\!=\!\pi R^2/\Delta d\,$ is the CN volumetric fraction in the planar dielectric film as sketched in Fig.~\ref{fig1}. The analytic continuation into the real frequency domain of our interest here can now be obtained by changing $i\omega$ to $\omega+i\delta$ in the above, to finally result in
\begin{equation}
\frac{\varepsilon_{yy}(q,\omega)}{\epsilon}\!=\!1\!-\!\frac{2f_{\rm CN\,}\sigma_{yy}(q,\omega)}{f_{\rm CN\,}\sigma_{yy}(q,\omega)+i\omega e^2N_{\rm 2D}R/m^\ast\omega^2_p(q)d}
\label{epsilonyy}
\end{equation}
with the longitudinal conductivity of an isolated SWCN to include both intra- and interband contributions, taking per Eqs.~(\ref{redef}), (\ref{sigmaintra}) and (\ref{Eq:sigma:Ando}) the form
\begin{eqnarray}
\sigma_{yy}(q,\omega)\!=\!\sigma_{yy}^{intra}\!+\sigma_{yy}^{inter}\!=i\frac{e^2N_{\rm 2D}}{m^\ast}\frac{\omega+i\delta}{(\omega+i\delta)^2-(v_Fq)^2}\nonumber\\
+\frac{\hbar e^2}{2\pi RL}\sum_s\frac{-2i\hbar\omega|\langle s|\hat{v}_y|0\rangle|^2}{E_s\big[E_s^2-(\hbar\omega)^2-2i\hbar^2\omega\delta\big]}\,.\hskip1.5cm
\label{sigmayy}
\end{eqnarray}
Here, $\delta$ takes the meaning of a (phenomenological) inverse energy relaxation time, $\delta\!=\!1/\tau_r$, associated with the inelastic (phonon, defect) scattering~\cite{BoWoTa2009,Perebeinos07} and can generally be different for intra- and interband processes.

Under continuous low-intensity light illumination, the dielectric film with the SWCN array embedded in it can still be treated as being at the thermal equilibrium. At not too low temperatures the distribution of the $s$-subband quasiparticle excitations over the $q$-momentum space is then given by the normalized Boltzmann distribution function
\begin{equation}
f_s(q,T)=\frac{1}{Q_s}\,e^{-\beta\hbar\omega_s(q)}
\label{Bfunc}
\end{equation}
with $\hbar\omega_s(q)$ representing the energy of the excited quasiparticle \emph{eigen} states of the SWCN array given by Eq.~(\ref{eigenen}) and the normalization factor is
\[
Q_s=\sum_{q}e^{-\beta\hbar\omega_s(q)}=\frac{L}{\pi\hbar}\sqrt{\!\frac{M_s}{2}}\!\int_{\!E_{exc}}^\infty\!\!\frac{dE\,e^{-\beta\hbar\omega_s(q)}}{\sqrt{E-E_{exc}(s)}}
\]
with $q=\!\sqrt{2M_s\,[E-\!E_{exc}(s)]}/\hbar$ to be used in the integral. The thermal averaging of Eq.~(\ref{epsilonyy}) with this distribution function gives the temperature dependence of the EM response of the SWCN array in the form
\begin{eqnarray}
\varepsilon_{yy}(T,\omega)=\sum_{q}f_s(q,T)\,\varepsilon_{yy}(q,\omega)\hskip0.3cm\label{epsilonyyT}\\
=\frac{L}{\pi\hbar}\sqrt{\!\frac{M_s}{2}}\!\int_{\!E_{exc}}^\infty\!\!\!\frac{dE\,f_s(q,T)\,\varepsilon_{yy}(q,\omega)}{\sqrt{E-E_{exc}(s)}}\,.\nonumber
\end{eqnarray}

Equations~(\ref{epsilonyy})-(\ref{epsilonyyT}) and (\ref{eigenen}) provide the complete description of the spatially anisotropic linear EM response of the periodically aligned SWCN array in the TD regime. This is generally a two-component spatially dispersive (nonlocal) tensor of the form
\begin{equation}
\hat{\varepsilon}(q,\omega)=\left[\begin{array}{cc}\varepsilon_{xx}&0\\0&\varepsilon_{yy}\end{array}\right]
=\left[\begin{array}{cc}\epsilon&0\\0&\varepsilon_{yy}(q,\omega)\end{array}\right],
\label{dieltensor}
\end{equation}
which should be averaged per Eq.~(\ref{epsilonyyT}) to represent the thickness-dependent anisotropic dynamical response to continuous low-intensity light illumination. In practice, the CN-array-embedding dielectric layer may still have inclusions of non-identical parallel aligned CNs. Assuming their (quasi-)periodic in-plane distribution, these inhomogeneities in an otherwise homogeneous SWCN array can be accounted for by means of the Maxwell-Garnett (MG) mixing method~\cite{Mar2016}, whereby under continuous illumination the $yy$-component of Eq.~(\ref{dieltensor}) takes the form
\begin{equation}
\left<\varepsilon_{yy}(T,\omega)\right>=\sum_{\alpha=1}^{n}w_\alpha\varepsilon_{yy}^{(\alpha)}(T,\omega)\,.
\label{MG:Mixing}
\end{equation}
Here, $\varepsilon_{yy}^{(\alpha)}(T,\omega)$ refers to a thermally averaged homogeneous component of weight
\begin{equation}
w_\alpha=\frac{f^{(\alpha)}_{\rm CN}}{\sum_{i=1}^nf^{(i)}_{\rm CN}}=\left[\sum_{i=1}^n\!\left(\frac{R_i}{R_\alpha}\right)^{\!\!2}\!\frac{\Delta_\alpha}{\Delta_i}\right]^{-1}
\label{MGweight}
\end{equation}
in the generally $n$-component inhomogeneous mixture of periodic (or quasiperiodic) SWCN arrays with radii $R_\alpha$ and intertube separation distances $\Delta_\alpha$, which is embedded in the finite-thickness dielectric layer. The weights are defined to give $\sum_{\alpha=1}^{n}w_\alpha\!=\!1$, accordingly.

\section{Discussion}\label{Sec:4}

One can see from Eq.~(\ref{eigenen}) that the intertube dipole-dipole interaction (\ref{Eq:Hint2q}) makes the CN array collective spatial dispersion $\hbar\omega_s(q)$ greatly different from the isolated-CN spatial dispersion $E_s(q)$ in Eq.~(\ref{Ef}) due to the nonzero ratio $V_{ss}(q)/E_s(q)$. This is a $q$-dependent function inversely proportional to $\Delta\!\in\!(2R,+\infty)$, which only tends to zero to give $\hbar\omega_s(q)\!=\!E_s(q)$ in the limit of $\Delta\!\rightarrow\!+\infty$. For small momenta $q$ controlling the most of the $q$-space quasiparticle state population, the major $q$-dependence comes from the linear dependence of $\omega_p^2(q)\!\sim\!q$ in $V_{ss}(q)$ [see Eq.~(\ref{plasmaFy})] rather than from the quadratic dependence of $E_s(q)\!\sim\!q^2$, which is why [as well as in view of the fact that $\hbar^2q^2/2M_{ex}\!\ll\!E_{exc}$ in Eq.~(\ref{Ef})] the second-order $q^2$-dispersion can be ignored for the qualitative analysis. With this in mind, comparing Eqs.~(\ref{Eq:Hint2q}) and (\ref{sigmayy}) with Eq.~(\ref{veldisp}) taken into account, one obtains an estimate
\begin{equation}
V_{ss}(q)\approx\frac{m^{\ast}\hbar^2\omega_p^2(q)}{e^2N_{\rm 2D}\tau_rE_s}\frac{2\pi R}{\Delta}\,\sigma_{yy}^{inter}(E_s),
\label{Vssapprox}
\end{equation}
where $E_s\!=\!E_{exc}$ and $\sigma_{yy}^{inter}(E_s)$ is the $\sigma_{yy}^{inter}(0,\omega\!=\!E_s/\hbar)$ single-subband approximation given by the greatest resonance term of the sum in Eq.~(\ref{sigmayy}). Substituting $\omega_p(q)$ of Eq.~(\ref{plasmaFy}) in Eq.~(\ref{Vssapprox}), one gets the ratio
\begin{equation}
\frac{V_{ss}(q)}{E_s}\approx g(E_s,\tau_r)\frac{2(qR)^2I_0(qR)K_0(qR)}{\epsilon qd+\epsilon_1+\epsilon_2}\frac{2\pi R}{\Delta}
\label{VsEsratio}
\end{equation}
with $g(E_s,\tau_r)\!=\!4\pi\hbar^2\sigma_{yy}^{inter}(E_s)/\tau_rE_s^2R$ being a constant.

\begin{figure}[t]
\includegraphics[width=1.0\linewidth]{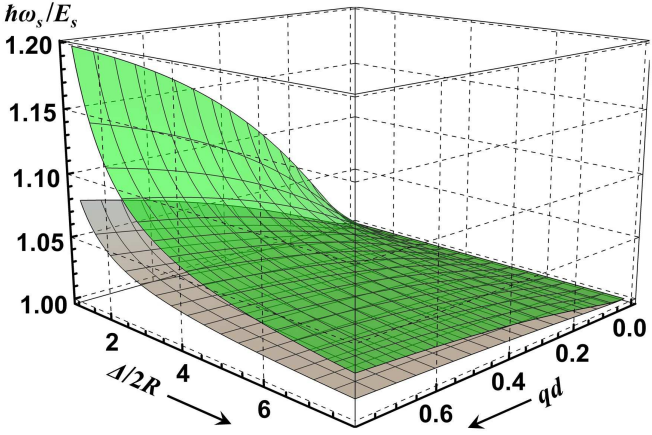}
\caption{Dispersion of collective excitations in the ultrathin SWCN arrays with $R/d\!=\!0.5$ (green) and $0.25$ (gray) relative to the isolated SWCN exciton dispersion as a function of array parameters $qd$ and $\Delta/2R$.}
\label{fig2}
\end{figure}

Plugging Eq.~(\ref{VsEsratio}) in Eq.~(\ref{eigenen}) gives the graph in Fig.~\ref{fig2} to show the general behavior of $\hbar\omega_s(q)/E_s$ as a function of $qd$ and $\Delta/2R$ for $R/d=0.5$ and $0.25$, the two cases where the array is enclosed in and buried inside of the dielectric layer, respectively. We used $E_s\!=\!E_{exc}\!\approx\!1$~eV and $\sigma_{yy}^{inter}(E_s)\!\approx\!100\,(e^2\!/2\pi\hbar)$ typical of the first-subband exciton in semiconducting SWCNs of $R\!\approx\!1$~nm~\cite{An2005,BoMe2014}, with $\tau_r\!=\!100$~fs typical of exciton-phonon scattering~\cite{Perebeinos07}, to obtain $g(E_s,\tau_r)\approx2$. The dielectric constants were set to $\epsilon\!=\!10$ and $\epsilon_{1,2}\!=\!1$. One can see a rapidly increasing spatial dispersion as the intertube distance shrinks. As expected, the thinner the dielectric layer ($d\!=\!2R$ compared to $d\!=\!4R$), the stronger the dispersion due to the dielectric screening effect reduction.

To better understand the dielectric response behavior as a function of the intrinsic parameters of the SWCN array, we rewrite Eq.~(\ref{epsilonyy}) as follows
\begin{equation}
\frac{\varepsilon_{yy}(q,\omega)}{\epsilon}=1-\frac{2[1+\lambda(\omega)]}{1+\lambda(\omega)+\omega(\omega+i\delta)\Delta/\omega^2_p(q)\pi R}\,.
\label{epslambda}
\end{equation}
Here, the array parameters are grouped in the denominator second term, the ratio
\begin{equation}
\lambda(\omega)\!=\!\frac{\sigma_{yy}^{inter}(\omega)}{\sigma_{yy}^{intra}(\omega)}\approx\!\sum_s\frac{\sigma_{yy}^{inter}(E_s)}{\sigma_{yy}^{intra}(0)}
\frac{2\hbar\omega(\hbar\omega+i\hbar\delta)}{(\hbar\omega)^2\!-E_s^2\!+2i\hbar^2\omega\delta}
\label{lambda}
\end{equation}
where in view of Eq.~(\ref{Vssapprox}) one has
\begin{equation}
\frac{\sigma_{yy}^{inter}(E_s)}{\sigma_{yy}^{intra}(0)}\approx\frac{E_sV_{ss}(q)}{\hbar^2\omega_p^2(q)}\frac{\Delta}{2\pi R}\,,
\label{sigEsig0}
\end{equation}
represents the SWCN alone, and the $q^2$-dispersion terms are dropped. The function $\lambda(\omega)$ can be seen to rapidly diminish for $\omega\!\ll\!E_s/\hbar$, whereby in the low-frequency range $\omega_p\sqrt{\pi R/\Delta}\!<\!\omega\!<\!E_s/\hbar$ far from resonances Eq.~(\ref{epslambda}) takes a spatially dispersive (non\-local) Drude-like response form
\begin{equation}
\frac{\varepsilon_{yy}(q,\omega)}{\epsilon}\approx1-\frac{2\pi R}{\Delta}\frac{\omega^2_p(q)}{\omega(\omega+i\delta)}\,,
\label{DrudeLimit}
\end{equation}
thickness/substrate/superstrate-dependent per Eq.~(\ref{plasmaFy}), which (except for a geometry-specific constant prefactor) was previously reported to consistently describe the optical properties of both in-plane isotropic and cylindrically anisotropic TD plasmonic films~\cite{BoMoSh2020,Bo2019,BoMoSh2018,BoSh2017,VeEtal2019}.

For $\hbar\omega\!\sim\!E_s$, the function $\lambda(\omega)$ can be seen to increase dramatically, whereby Eq.~(\ref{epslambda}) after straightforward simplifications using Eq.~(\ref{sigEsig0}) takes the form
\[
\frac{\varepsilon_{yy}(q,\omega)}{\epsilon}\approx1\!-\!\frac{2E_sV_{ss}(q)}{(\hbar\omega)^2\!-\!\big[1\!-\!V_{ss}(q)/E_s\big]E_s^2\!+\!2i\hbar^2\omega\delta}\,,
\]
which can further be compacted to give
\begin{eqnarray}
\frac{\varepsilon_{yy}(q,\omega)}{\epsilon}\approx\frac{\big[(\hbar\omega)^2\!-\!E_{s+}^2\big]\big[(\hbar\omega)^2\!-\!E_{s-}^2\big]}{\big[(\hbar\omega)^2\!-\!E_{s-}^2\big]^2\!+\big(2\hbar^2\omega\delta\big)^2}
\label{epsResEs}\\
+\,i\frac{4E_sV_{ss}(q)\hbar^2\omega\delta}{\big[(\hbar\omega)^2\!-\!E_{s-}^2\big]^2\!\!+\big(2\hbar^2\omega\delta\big)^2}\,,\hskip0.75cm\nonumber
\end{eqnarray}
where $E_{s\pm}\!=\!E_s\sqrt{1\pm V_{ss}(q)/E_s}\,$. Here, both absorption resonance position and intensity of the EM response can be seen being adjustable by varying the array thickness, density and dielectric parameters according to Eq.~(\ref{VsEsratio}), whereby the ratio $V_{ss}(q)/E_{s\!}=\!\{[\hbar\omega_s(q)/E_s]^2-1\}/2$ can be adjusted that controls the difference between the collective eigen-state and single-tube exciton-state energies. The real part $\mbox{Re}\,\varepsilon_{yy}/\epsilon$ of Eq.~(\ref{epsResEs}) reveals the \emph{negative refraction} (NR) band centered around $E_s$ in the domain
\begin{equation}
\sqrt{E_s^2\!-\!E_sV_{ss}(q)}\!=\!E_{s-\!}\!<\!\hbar\omega\!<\!E_{s+\!}\!=\!\sqrt{E_s^2\!+\!E_sV_{ss}(q)}
\label{negativeband}
\end{equation}
of width $E_{s+\!}-E_{s-}\!\approx\!V_{ss}(q)$. The imaginary part $\mbox{Im}\,\varepsilon_{yy}/\epsilon$ of Eq.~(\ref{epsResEs}) and the imaginary part of its negative inverse
\begin{eqnarray}
-\mbox{Im}\frac{\epsilon}{\varepsilon_{yy}(q,\omega)}\approx\hskip2.75cm
\label{Im1overeps}\\
\frac{4E_sV_{ss}(q)\hbar^2\omega\delta\,\Big\{\!\big[(\hbar\omega)^2\!-\!E_{s-}^2\big]^2\!+\big(2\hbar^2\omega\delta\big)^2\!\Big\}}
{\Big\{\!\big[(\hbar\omega)^2\!-\!E_{s+}^2\big]\big[(\hbar\omega)^2\!-\!E_{s-}^2\big]\!\Big\}^2\!\!+\big[4E_sV_{ss}(q)\hbar^2\omega\delta\big]^2}\nonumber
\end{eqnarray}
are representative of the transversely polarized EM wave absorption by excitons and of the Coulomb-energy losses for longitudinally polarized plasmon excitations, respectively~\cite{Cardona}. From Eqs.~(\ref{epsResEs}) and (\ref{Im1overeps}) $E_{s-}$ and $E_{s+}$ can be seen being their respective resonance peak energies. The latter is an analogue of the \emph{interband} plasmon resonance, the one situated in-between the two neighboring exciton subbands~\cite{BoWoTa2009,Bo2012}).

The range of the NR band can be found straightforwardly from the analysis of extrema for the real-valued function $\mbox{Re}\,\varepsilon_{yy}/\epsilon$ of Eq.~(\ref{epsResEs}). This gives
\begin{equation}
\big(\hbar\omega_{min}\big)^2\!\approx E_{s-}^2\!+2E_{s}\hbar\delta\sqrt{1\!+\!\Big[\frac{\hbar\delta}{V_{ss}(q)}\Big]^{2}}
\label{omegamin}
\end{equation}
to the first infinitesimal order in the smallness parameter $\hbar\delta/E_s\!\ll1$ and
\begin{equation}
\mbox{Re}\,\frac{\varepsilon_{yy}(q,\omega_{min})}{\epsilon}\approx\frac{1}{2}\Big\{1-\sqrt{1\!+\!\Big[\frac{V_{ss}(q)}{\hbar\delta}\Big]^{2}}\Big\}
\label{reepsmin}
\end{equation}
to the first \emph{nonvanishing} (zeroth) order in the same parameter. In these equations the ratio under the square root is independent of $\delta$ since $V_{ss}(q)\!\sim\!\hbar/\tau_r\!=\hbar\delta$ as can be seen from Eq.~(\ref{VsEsratio}). This ratio can be controlled by means of $\Delta/2R$ to rarefy (or densify) the CN array as well as through $d$ and $\epsilon/(\epsilon_1+\epsilon_2)$ to adjust the array thickness and material composition, respectively. From Eqs.~(\ref{negativeband}), (\ref{omegamin}) and (\ref{reepsmin}) it can be seen that $\hbar\omega_{min}$ shifts to the blue and the NR bandwidth shrinks down to zero along with its range as $V_{ss}(q)$ decreases. Clearly, originating from the intertube dipole-dipole interaction, the NR band of the homo\-geneous TD arrays of identical SWCNs can be controlled on demand, both to expand and to shrink, by means of adjusting these interactions per Eq.~(\ref{VsEsratio}).

\begin{figure}[t]
\hskip0.5cm\includegraphics[width=.82\linewidth]{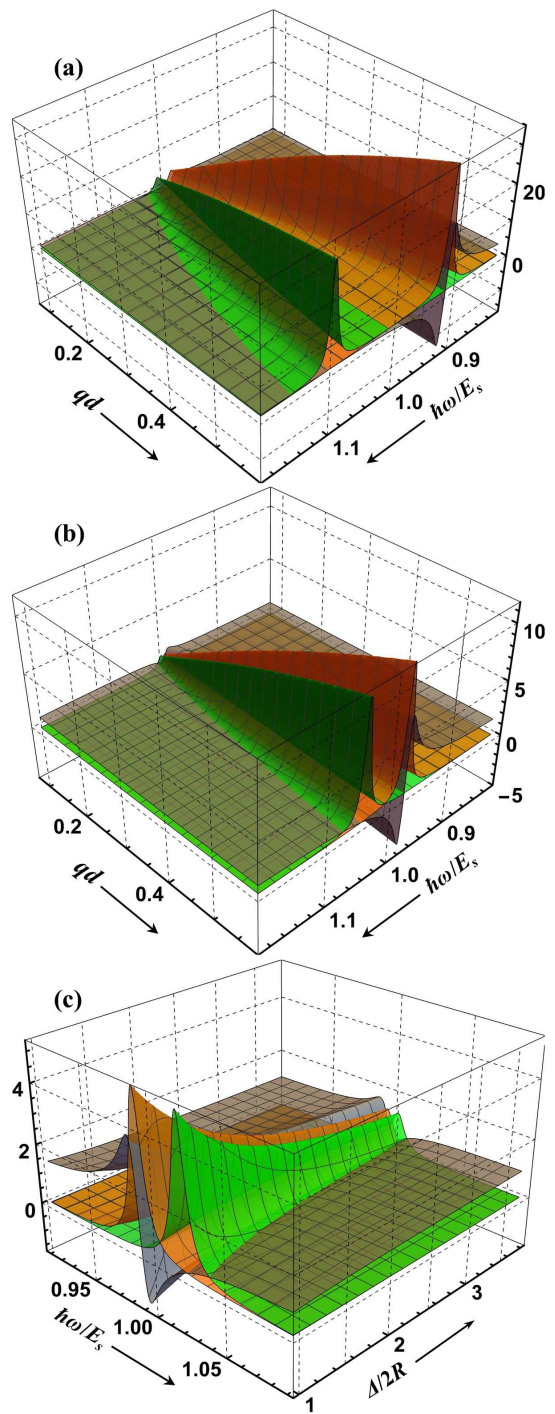}
\caption{$\mbox{Re}\,\varepsilon_{yy}/\epsilon$ (gray), $\mbox{Im}\,\varepsilon_{yy}/\epsilon$ (orange) and $-\mbox{Im}(\epsilon/\varepsilon_{yy})$ (green) plotted per Eq.~(\ref{epsResEs}) in the exciton resonance range $\hbar\omega/E_s\!\sim\!1$ as functions of $\hbar\omega/E_s$ and $qd$ for dense ($\Delta/2R\!=\!1$) ultra\-thin SWCN arrays with $R/d\!=\!0.5$ (a) and $0.25$ (b), and as functions of $\hbar\omega/E_s$ and $\Delta/2R$ with $R/d\!=\!0.5$, $qd\!=\!0.1$~(c).}\vspace{-0.57cm}
\label{fig3}
\end{figure}

Figure~\ref{fig3} shows $\mbox{Re}\,\varepsilon_{yy}/\epsilon$, $\mbox{Im}\,\varepsilon_{yy}/\epsilon$ and $-\mbox{Im}(\epsilon/\varepsilon_{yy})$ in dimensionless variables as given by Eqs.~(\ref{epsResEs}), (\ref{Im1overeps}) and (\ref{VsEsratio}), to demonstrate the NR band behavior, the exciton absorption and interband plasmon response in the neighborhood of a single-tube exciton resonance as functions of the intrinsic parameters of the SWCN array. The graphs are obtained for the same material parameters as in Fig.~\ref{fig2}.~In~(a) and (b), the decrease of the thickness can be seen to push the exciton and plasmon resonances closer together, quenching their intensities to reduce the NR bandwidth and range. Comparing (a) and (b), both exciton and plasmon resonance intensities can be seen being lower for the thicker dielectric layer in (b), apparently due to the greater dielectric screening effect. In (c), the overlap of the exciton and plasmon resonances can be seen to increase with the intertube distance, whereby a controllable exciton-plasmon \emph{hybridization} is possible by varying the intrinsic parameters of the array. All these are the \emph{universal} features of the EM response that originate from the eigenstate dispersion relations of the TD arrays of SWCNs in the domain of their interband transitions as described by Eqs.~(\ref{epsResEs})-(\ref{reepsmin}). Far from this domain their low-energy EM response features a thickness/substrate/superstrate-dependent Drude-like metallic behavior given by Eq.~(\ref{DrudeLimit}).

\begin{figure}[t]
\hskip-0.3cm\includegraphics[width=1.0\linewidth]{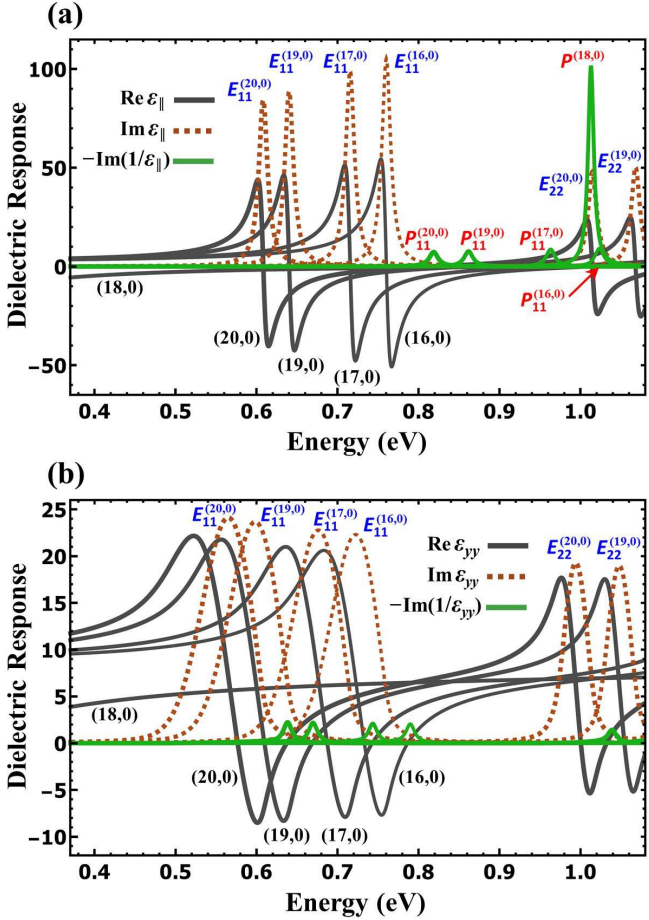}
\caption{(a)~The individual dielectric responses along the CN axis (longitudinal) for the zigzag (16,0), (17,0), (18,0), (19,0), and (20,0) SWCNs \emph{in vacuum}, obtained using the $\textbf{k}\cdot\textbf{p}$ method of the SWCN band structure calculations~\cite{An2005}. All graphs are scaled down vertically by a factor of $10$ for better visibility. (b)~The respective room-temperature dielectric response functions along the CN alignment direction, calculated as given by Eqs.~(\ref{epsilonyy})-(\ref{epsilonyyT}) for the ultrathin periodic arrays of the (16,0), (17,0), (18,0), (19,0) and (20,0) SWCNs. See text for details.}
\label{fig4}
\end{figure}

\begin{figure}[t]
\hskip-0.3cm\includegraphics[width=1.03\linewidth]{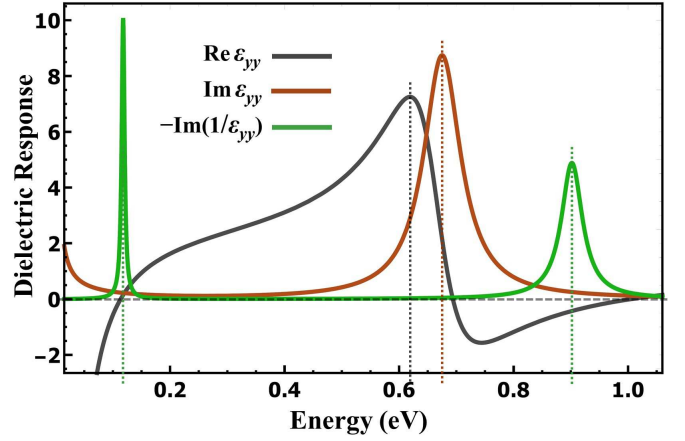}
\caption{The room-temperature ($300$~K) in-plane dielectric response functions along the CN alignment direction calculated for an ultrathin ($\sim\!10$~nm) weakly inhomogeneous TD film of the MG-mixed (16,0), (17,0), (18,0), (19,0) and (20,0) SWCN arrays whose individual responses are presented in Fig.~\ref{fig4}~(b). See text for details.}
\label{fig5}
\end{figure}

Figures~\ref{fig4} and~\ref{fig5} present our numerical simulations to show how the thermal broadening and a slight diameter dispersion affect the EM response of a finite-thickness SWCN film in the TD regime. We start with the longitudinal (along the CN symmetry axis) dielectric functions $\varepsilon_\parallel(\omega)$ for the five individual zigzag SWCNs in vacuum, shown in Fig.~\ref{fig4}~(a). The SWCNs chosen are the semiconducting (16,0), (17,0), (19,0) and (20,0) CNs and the metallic (18,0) CN --- all of about the same diameter centered at $2R_{(18,0)}\!=\!1.41$~nm. Focusing on the domain around the first exciton resonance, we obtain their dielectric functions $\varepsilon_\parallel$ from $\sigma^{inter}_{yy}$ of Eq.~(\ref{sigmayy}), calculated using the $\textbf{k}\!\cdot\!\textbf{p}$ method of the SWCN band structure calculations~\cite{An2005} with $\tau_r\!=\!100$~fs, followed by the Drude relation
$\varepsilon_{\parallel}(\omega)\!=1\!+8\pi i\sigma_{\parallel}(\omega)/R\omega$~\cite{BoWoTa2009} with $\sigma_{\parallel}\!=\sigma_{yy}$ of Eq.~(\ref{sigmayy}).

Marked in Fig.~\ref{fig4}~(a) are peaks of $\mbox{Im}\varepsilon_{\parallel}$ and $-\mbox{Im}(1/\varepsilon_{\parallel})$ to indicate the dipole \emph{interband} electronic transitions $E_{11}$ ($E_{22}$) and $P_{11}$ from the CN first (second) valence band to the CN first (second) conduction band~\cite{An2005}, to produce the first (second) exciton and the first interband plasmon, respectively. The latter one, though not that intensive, shows up in the domains where $\mbox{Re}\varepsilon_{\parallel\!}\!=\!0$ and $\mbox{Im}\varepsilon_{\parallel\!}\!\rightarrow\!0$ (or $\mbox{Im}\sigma_{\parallel\!}\!=\!0$ and $\mbox{Re}\sigma_{\parallel\!}\!\rightarrow\!0$, accordingly~\cite{BoWoTa2009,Bo2012}) due to the Kramers-Kronig relation that links $\mbox{Re}\varepsilon_{\parallel}$ and $\mbox{Im}\varepsilon_{\parallel}$. The plasmon peak $P^{(18,0)}$ of the metallic (18,0) SWCN is due to the \emph{intraband} transition that comes from $\sigma^{intra}_{yy}$ of Eq.~(\ref{sigmayy}). As a classical plasmon resonance it is highly pronounced, while the interband (quantum) transitions occur for the (18,0) SWCN at much higher energies outside of the domain presented. For each of the nanotubes in Fig.~\ref{fig4}~(a), their respective dielectric response functions exhibit sharp resonance structures typical of vacuum and known to get quenched by a dielectric background~\cite{Ando2010}. To obtain the dielectric responses for the respective homogeneous SWCN arrays, presented in Fig~\ref{fig4}~(b) in the same energy domain for comparison, we use our derived Eqs.~(\ref{epsilonyy})-(\ref{epsilonyyT}) with $d\!=\!\Delta\!=\!2R$ yielding $f_{\rm CN}\!=\!\pi/4$ for their respective $R$, and $T\!=\!300$~K for each of the arrays. Other array parameters are taken to be the same for all five of them: $\epsilon\!=\!10$, $\epsilon_{1}\!=\!\epsilon_{2}\!=\!1$, the first exciton translational effective mass $M_{1\!}\!=\!0.4\,m_0$ ($m_0$ is the free electron rest mass) and the relaxation time $\tau_r\!=\!100$~fs~\cite{BoWoTa2009,Perebeinos07}. The intensity decrease and large broadening can be seen along with the red shifts of the exciton and plasmon resonances, overall just in the way prescribed universally by the single-resonance approximation as per Eqs.~(\ref{epsResEs})-(\ref{reepsmin}), including the presence of the NR band.

Figure~\ref{fig5} presents an example of the in-plane dielectric responses along the CN alignment direction for a weakly inhomogeneous TD film made out of a quasi\-periodic mixture of the (16,0), (17,0), (18,0), (19,0) and (20,0) homo\-geneous SWCN arrays.~To obtain these graphs, we use the MG method as prescribed in Eqs.~(\ref{MG:Mixing}) and (\ref{MGweight}). For each array component in the mixture the thickness is taken to be $d\!=\!3R$ and the intertube distances $\Delta_{\alpha}\,(\alpha\!=\!1,...,5)$ are set to $\Delta_{(16,0)\!}\!=\!6R$, $\Delta_{(17,0)\!}\!=\!2R$, $\Delta_{(18,0)\!}\!=\!2R$, $\Delta_{(19,0)\!}\!=\!5R$ and $\Delta_{(20,0)\!}\!=\!6R$ with $R$ being the radius of the constituent SWCN of the respective homo\-geneous array. The resulting weakly inhomogeneous TD film comes out to consist of the (16,0), (17,0), (18,0), (19,0) and (20,0) homogeneous SWCN arrays with the relative weights of 0.1, 0.31, 0.33, 0.14 and 0.12, respectively, to give the sum of weights equal to unity as required and to make the film be composed of about $1/3$ metallic and $2/3$ semiconducting SWCNs which is normally the case experimentally~\cite{RoEtal2019,SC2020,JFan2020}. The dielectric responses of the individual array components are calculated beforehand as described for Fig.~\ref{fig4}~(b) above. The overall thickness of this film is estimated to be $\sim\!10$~nm. We also take into that in a real sample there may be other CNs of different chiralities but of about the same diameter, some of them may be bent, tilted and so may not be equally spaced or precisely periodic. Therefore, the curves in Fig.~\ref{fig5} are smoothed out to a regular shape.

Comparing Fig.~\ref{fig5} and Fig~\ref{fig4}~(b), one can see a much stronger broadening for both exciton and interband plasmon resonances of the (just weakly) inhomogeneous film. In contrast to its homogeneous SWCN array constituents, the broaden interband plasmon resonance of the MG-mixed film is almost as intensive as the exciton resonance. Both resonances can now be seen to overlap a great deal, thus making possible the exciton-plasmon coupling and the respective exciton-plasmon hybridization. We believe this sheds more light on the nature of the ultrastrong exciton-plasmon coupling effect recently reported for self-assembled crystallized CN films experimentally~\cite{HoEtAl2018}. The NR band of the weakly inhomogeneous TD film in Fig.~\ref{fig5} can be seen to expand and its range to shrink accordingly, breaking the limits set up for homogeneous periodic SWCN arrays by Eqs.~(\ref{negativeband}) and (\ref{reepsmin}). The classical intraband plasmon resonance, a low-energy feature of the metallic (18,0) array component described by Eq.~(\ref{DrudeLimit}) and not shown in Fig.~\ref{fig4}~(b), is presented in Fig.~\ref{fig5} as well. Since we only have a single metallic CN component in our model MG-mixture here, this resonance comes out very narrow in the graphs presented. However, for instance, having the metallic (16,1) and (19,1) SWCNs included, with diameters of just $8\,\%$ less and greater than that of the (18,0) SWCN, respectively, would broaden this resonance inhomogeneously as well. Furthermore, in addition to the phonon scattering, intraband plasma oscillations can be strongly quenched by the Coulomb electron scattering, a scattering process that is suppressed for interband transitions forming neutral excitons, thus to result in a much stronger homogeneous broadening of the intraband plasmon resonance. In real self-assembled CN films one therefore should expect this resonance to be largely broaden, both homogeneously and inhomogeneously.

\section{Conclusions}\label{Sec:5}

In this paper, we study the intrinsic collective quasiparticle excitations responsible for the in-plane EM response of the ultrathin plane-parallel homogeneous periodic SWCN arrays and weakly inhomogeneous SWCN films in the TD regime.~We use the low-energy plasmon response calculation technique~\cite{Bo2019} combined with the many-particle Green's function formalism in the Matsubara formulation~\cite{Mahan2000} to derive the dynamical dielectric tensor of the system in the broad spectral domain of microwave to visible ($\lesssim\!1\!-\!2$~eV). This is where intrinsic excitations, excitons and plasmons, are present in individual constituent SWCNs~\cite{Bo15,DrSaJo2007}. We use an approach that works well as applied to molecular solids where individual molecules, represented by individual nano\-tubes in our case, are weakly bound together by the van der Waals interaction, and the EM response of the solid (a dense periodic array or film of aligned nanotubes in our case) comes about due to a locally induced single-molecule dipole polarization propagating through the periodic lattice of molecules to create the collective polarization of the entire molecular solid. Our theory links the dynamical dielectric response tensor of the SWCN periodic array to the complex longitudinal conductivity of the constituent SWCN, a building block the array is composed of. The SWCN conductivity can be calculated by a number of numerical and analytical methods~\cite{An2005,DrSaJo2007,Ando2009}, of which to demonstrate the way our theory works we use the $\textbf{k}\!\cdot\!\textbf{p}$ method of the SWCN band structure calculations based on the Kubo linear-response theory~\cite{An2005}. Our model and the results we present encompass the periodic arrays and films formed by achiral SWCNs. Arrays of chiral SWCNs require an extra analysis for collective gyrotropic effects, which we will present separately elsewhere.

We show that the collective dielectric response can be controlled by the volume fraction of constituent SWCNs, the active component of the ultrathin TD film, and that this can be done not only by varying the parameters of the SWCN content, such as the CN diameter and intertube distance, but also by merely varying the thickness of the CN embedding dielectric layer. For homogeneous single-type SWCN periodic arrays, the real part of the dielectric response function is negative for a sufficiently wide domain in the neighborhood of a quantum interband transition of the constituent SWCN, to form a relatively broad NR band that makes the system behave as a hyper\-bolic metamaterial at much higher frequencies than those typically in the IR domain that the classical intraband plasma oscillations have to offer~\cite{Chen2020}. By decreasing the CN diameter it is quite possible to push this NR band in the visible region, while using weakly inhomogeneous multi-type SWCN films can make it even broader than that of the single-type SWCN array.

The overall behavior we obtain theoretically for the real and imaginary part of the in-plane dielectric response of the weakly inhomogeneous SWCN film is very similar to that reported experimentally for self-assembled SWCN metamaterial films in Ref.~\cite{RoEtal2019}, which is why we believe that the NR band observed there for undoped samples is most likely due to the inhomogeneous broadening as presented in Fig.~\ref{fig5}. Electron doping would primarily affect the intraband plasmon of a metallic CN component to largely broaden and shift it to the blue, as evidenced by our Eq.~(\ref{DrudeLimit}), due to the Coulomb scattering and electron density increase, respectively, which is precisely what was observed in Ref.~\cite{RoEtal2019} as well.~Our theory thereby indicates that the periodically aligned, homogeneous and weakly inhomogeneous ultrathin TD films of SWCNs are excellent candidates for the development of a new generation of advanced multifunctional optical hyperbolic metasurfaces, to push the NR band and the respective hyperbolic response (typically pertinent to the IR~\cite{Chen2020}) into the optical spectral region, with characteristics adjustable on demand by means of the SWCN diameter, chirality and periodicity variation.

\section*{Acknowledgments}

This research is supported by the U.S. National Science Foundation under Condensed Matter Theory Program Award No. DMR-1830874 (I.V.B.)

\appendix

\onecolumngrid

\section{~Proof of Eq.~(\ref{Eq:Fyy0:Simp})}\label{Appx1}

Under constant illumination by low-intensity external monochromatic radiation polarized along the nanotube alignment direction, by virtue of the linear response theory in the exciton-radiation interaction and associated fluctuation-dissipation theorem~\cite{LaLi1980}, the averaging over all possible SWCN array quasiparticle configurations can be done in the form of a general equilibrium statistical ensemble averaging as follows
\begin{equation}
\langle\,\cdots\rangle=e^{\beta\Omega}\,\mbox{Tr}\big(e^{-\beta\hat{H}}\cdots\big)~~\mbox{with}~~e^{-\beta\Omega}\!=\mbox{Tr}\big(e^{-\beta\hat{H}}\big),
\label{thermalavdef}
\end{equation}
where $\hat{H}\!=\hat{H}_0+\hat{H}_{int}$ is the total Hamiltonian~(\ref{Eq:Htotfin}) of the system. For the reasons explained in the main text we choose not to use the diagonalized form (\ref{Eq:Htotdiag}) of the total Hamiltonian here. Instead, we proceed in the standard way~\cite{Mahan2000,Abr1975}, working in the noninteracting exciton Hilbert space
\begin{equation}
\prod_{s,q}|N_{s,q}\rangle=\prod_{s,q}\frac{\big(B_{s,q}^\dag\big)^{N_{s,q}}}{\sqrt{N_{s,q}!}}|0\rangle,
\label{HilbertSpace}
\end{equation}
with the noninteracting particle number operator defined by equations $\hat{N}_{s,q}|N_{s,q}\rangle\!=\!B_{s,q}^\dag B_{s,q}|N_{s,q}\rangle\!=\!N_{s,q}|N_{s,q}\rangle$, whereby the equilibrium statistical ensemble averaging procedure (\ref{thermalavdef}) takes the following explicit form
\begin{equation}
\langle\,\cdots\rangle=e^{\beta\Omega}\,\mbox{Tr}\big[e^{-\beta(\hat{H}_0+\hat{H}_{int})}\cdots\big]=e^{\beta\Omega}\prod_{s,q}\sum_{N_{s,q}=0}^\infty\langle N_{s,q}|\,e^{-\beta(\hat{H}_0+\hat{H}_{int})}\cdots|N_{s,q}\rangle.
\label{statav}
\end{equation}
In particular, in the absence of the intertube interaction $\hat{H}_{int}$ this takes the form
\begin{equation}
\langle\,\cdots\rangle_0=e^{\beta\Omega_0}\,\mbox{Tr}\big[e^{-\beta\hat{H}_0}\cdots\big]=e^{\beta\Omega_0}\prod_{s,q}\sum_{N_{s,q}=0}^\infty\langle N_{s,q}|\,e^{-\beta\hat{H}_0}\cdots|N_{s,q}\rangle
\label{statav01}
\end{equation}
with
\begin{equation}
e^{-\beta\Omega_0}=\mbox{Tr}\big(e^{-\beta\hat{H}_0}\big)=\prod_{s,q}\sum_{N_{s,q}=0}^\infty\langle N_{s,q}|\,e^{-\beta\sum_{s^\prime\!\!,q^\prime}\!E_{s^\prime}(q^\prime)\hat{N}_{s^\prime\!\!,q^\prime}}|N_{s,q}\rangle=\prod_{s,q}\frac{1}{1-e^{-\beta E_s(q)}}.
\label{statav02}
\end{equation}
This consistently gives the mean exciton occupation number of a noninteracting boson form
\begin{equation}
\bar{N}_{s,q}=\langle\hat{N}_{s,q}\rangle_0=\langle B_{s,q}^\dag B_{s,q}\rangle_0=\frac{1}{e^{\beta E_s(q)}-1}
\label{Nsqav}
\end{equation}
with no chemical potential in it since only one exciton can be thought of being present in the system at a time, which is typically the case for not too high irradiation intensity.

For the \emph{noninteracting} SWCN array the averaging in Eq.~(\ref{Eq:Fij0}) takes the form
\begin{equation}
\langle T_\tau\hat{d}_y(q,\tau)\hat{d}_y(-q,0)\rangle_0=\Theta(\tau)\langle\hat{d}_y(q,\tau)\hat{d}_y(-q,0)\rangle_0+\Theta(-\tau)\langle\hat{d}_y(-q,0)\hat{d}_y(q,\tau)\rangle_0,
\label{tauorder}
\end{equation}
where $\Theta(\tau)$ is the theta-function, to give
\begin{equation}
P_{yy}^{(0)}(q,i\omega)=-\frac{1}{\hbar}\int_{0}^{\hbar\beta}\!\!\!\!\!d\tau\,e^{i\omega\tau}\langle\hat{d}_y(q,\tau)\hat{d}_y(-q,0)\rangle_0
\label{Pyy0}
\end{equation}
with
\begin{equation}
\hat{d}_y(q,\tau)=e^{\tau\hat{H}_0}\hat{d}_y(q)e^{-\tau\hat{H}_0}
\label{dtau}
\end{equation}
and the (Schr\"{o}dinger-picture) induced dipole moment operator $\hat{d}_y(q)$ defined in terms of the exciton creation and annihilation operators per Eq.~(\ref{Eq:Hint1q}). Using the Baker-Hausdorff lemma (see, e.g., Ref.~\cite{AbersQM})
\[
e^{\hat{A}}\hat{C}e^{-\hat{A}}\!=\hat{C}+[\hat{A},\hat{C}]+\frac{1}{2!}[\hat{A},[\hat{A},\hat{C}]]+\frac{1}{3!}[\hat{A},[\hat{A},[\hat{A},\hat{C}]]]+\,\cdots
\]
it is easy to show that
\begin{equation}
e^{\tau\hat{H}_0}B_{s,q}e^{-\tau\hat{H}_0}=e^{-\tau E_s}B_{s,q}\,,\;\;\;e^{\tau\hat{H}_0}B^\dag_{s,q}e^{-\tau\hat{H}_0}=e^{\tau E_s}B^\dag_{s,q}\,.
\label{BtauEs}
\end{equation}
Inserting Eqs.~(\ref{dtau}) and (\ref{BtauEs}) in Eq.~(\ref{Pyy0}) leads to
\begin{equation}
P_{yy}^{(0)}(q,i\omega)=\!-\frac{m^\ast\omega^2_p(q)d}{4\pi\hbar N_{\rm 2D}N}\!\int_0^{\hbar\beta}\!\!\!\!\!d\tau\,e^{i\omega\tau}\!
\sum_s|\langle s|\hat{\beta}_y|0\rangle|^2\Big(e^{-E_s\tau/\hbar}\langle B_{s,q}B^\dagger_{s,q}\rangle_0+e^{E_s\tau/\hbar}\langle B^\dagger_{s,-q}B_{s,-q}\rangle_0\Big).
\label{Eq:Fyy0}
\end{equation}
This can be easily integrated to give
\begin{equation}
P_{yy}^{(0)}(q,i\omega)=-\frac{m^\ast\omega^2_p(q)d}{4\pi\hbar N_{\rm 2D} N}\sum_s|\langle s|\hat{\beta}_y|0\rangle|^2\Big(
\frac{e^{-\beta E_s}\!-1}{i\omega-E_s/\hbar}\langle B_{s,q}B^\dagger_{s,q}\rangle_0+\frac{e^{\beta E_s}\!-1}{i\omega+E_s/\hbar}\langle B^\dagger_{s,-q}B_{s,-q}\rangle_0\Big)
\label{Eq:Fyy0:AP}
\end{equation}
due to the constraint $i\omega\!=\!2n\pi/\hbar\beta$ ($n$ is an integer) on the Matsubara frequency for bosons~\cite{Mahan2000,Abr1975}, yielding $e^{i\omega\hbar\beta}\!=1$. From Eq.~(\ref{Eq:Fyy0:AP}), using Eq.~(\ref{Nsqav}), the boson commutation relations, and the fact that $E_s(q)$ is an even function of $q$, one finally arrives at Eq.~(\ref{Eq:Fyy0:Simp}).

\section{~Proof of Eq.~(\ref{Pyy})}\label{Appx2}

For the \emph{interacting} SWCN array, per Eq.~(\ref{statav}), the positive-$\tau$ statistical averaging in Eq.~(\ref{Eq:Fij0}) with the Heisenberg-picture dipole operators explicitly written is of the form
\[
e^{\beta\Omega}\,\mbox{Tr}\big[e^{-\beta(\hat{H}_0+\hat{H}_{int})}T_\tau e^{\tau(\hat{H}_0+\hat{H}_{int})}\hat{d}_y(q)e^{-\tau(\hat{H}_0+\hat{H}_{int})}\hat{d}_y(-q)\big],\;\;\;\;\;e^{-\beta\Omega}\!=\mbox{Tr}\big[e^{-\beta(\hat{H}_0+\hat{H}_{int})}\big].
\]
This can be equivalently recast (see Ref.~\cite{Mahan2000} for details) by writing all the operators in the interaction representation defined by Eq.~(\ref{dtau}), to take the form
\begin{equation}
e^{\beta\Omega}\,\mbox{Tr}\big[e^{-\beta\hat{H}_0}T_\tau S(\beta)\hat{d}_y(q,\tau)\hat{d}_y(-q,0)\big]
\label{StatAverage1}
\end{equation}
with
\begin{equation}
e^{-\beta\Omega}\!=\mbox{Tr}\big[e^{-\beta\hat{H}_0}S(\beta)\big],
\label{StatAverage2}
\end{equation}
where $S(\beta)$ stands for the $S$-matrix defined as follows
\begin{equation}
S(\beta)=T_\tau\exp\!\Big[-\!\frac{1}{\hbar}\int_0^{\hbar\beta}\!\!\!d\tau\,\hat{H}_{int}(\tau)\Big]=\sum_{n=0}^\infty \frac{(-1/\hbar)^n}{n!}\prod_{i,j=1}^n \int_0^{\hbar\beta}\!\!\!d\tau_i\,T_\tau\hat{H}_{int}(\tau_j).
\label{Sbeta}
\end{equation}

The $S$-matrix series expansion (\ref{Sbeta}) produces a series of summands in Eq.~(\ref{StatAverage1}), which after the $\tau$-ordered operator pairings have been done per Wick's theorem, can be classified in terms of the two types of Feynman diagrams. They are disconnected and connected diagrams (with and without "bubbles", respectively). More specifically~\cite{Mahan2000}, the $n\!\ge\!1$ terms of Eq.~(\ref{Sbeta}) generate in the exponential factor of Eq.~(\ref{StatAverage2}) the (vacuum polarization) disconnected diagrams of increasing order in $\hat{H}_{int}(\tau_j)$, which cancel out exactly the disconnected diagrams of the trace expression in Eq.~(\ref{StatAverage1}). As a result, Eq.~(\ref{StatAverage1}) takes the form
\begin{equation}
e^{\beta\Omega_0}\,\mbox{Tr}\big[e^{-\beta\hat{H}_0}T_\tau S(\beta)\hat{d}_y(q,\tau)\hat{d}_y(-q,0)\big]=\langle T_\tau S(\beta)\hat{d}_y(q,\tau)\hat{d}_y(-q,0)\rangle_0
\label{statav0}
\end{equation}
with $\langle\,\cdots\rangle_0$ defined by Eqs.~(\ref{statav01}) and (\ref{statav02}) where \emph{only} the connected diagrams are to be retained and summed up. The interacting SWCN array polarization of Eq.~(\ref{Eq:Fij0}) takes the form
\begin{equation}
P_{yy}(q,i\omega)=-\frac{1}{\hbar}\int_{0}^{\hbar\beta}\!\!\!\!\!d\tau\,e^{i\omega\tau}\langle T_\tau S(\beta)\hat{d}_y(q,\tau)\hat{d}_y(-q,0)\rangle_0,
\label{Pyynew}
\end{equation}
accordingly. Here, all the operators are written in the interaction representation defined by Eq.~(\ref{dtau}) and the averaging with the $S$-matrix of Eq.~(\ref{Sbeta}) expands explicitly as follows
\begin{equation}
\langle T_\tau S(\beta)\hat{d}_y(q,\tau)\hat{d}_y(-q,0)\rangle_0=\sum_{n=0}^\infty\frac{(-1/\hbar)^n}{n!}\!\!\int_0^{\hbar\beta}\!\!\!d\tau_1\cdots\!\!\int_0^{\hbar\beta}\!\!\!d\tau_n\,
\langle T_\tau \hat{d}_y(q,\tau)\hat{H}_{int}(\tau_1)\cdots\hat{H}_{int}(\tau_n)\hat{d}_y(-q,0)\rangle_0.
\label{Pyyexpansion}
\end{equation}
Putting in here the perturbation $\hat{H}_{int}$ in its explicit form (\ref{Eq:Hint1q}) brings the right-hand side of Eq.~(\ref{Pyyexpansion}) to the form
\begin{eqnarray}
\sum_{n=0}^\infty\frac{(-1/\hbar)^n}{n!}\!\!\int_0^{\hbar\beta}\!\!\!\!\!d\tau_1\cdots\!\!\int_0^{\hbar\beta}\!\!\!\!\!d\tau_n\,
\langle T_\tau\hat{d}_y(q,\tau)\frac{2\pi}{v_0}\sum_{q_1}\hat{d}_y(q_1,\tau_1)\hat{d}_y(-q_1,\tau_1)\cdots
\frac{2\pi}{v_0}\sum_{q_n}\hat{d}_y(q_n,\tau_n)\hat{d}_y(-q_n,\tau_n)\hat{d}_y(-q,0)\rangle_0,\nonumber
\end{eqnarray}
or more explicitly,
\begin{eqnarray}
\langle T_\tau\hat{d}_y(q,\tau)\hat{d}_y(-q,0)\rangle_0-\frac{2\pi}{\hbar v_0}\sum_{q_1}\int_0^{\hbar\beta}\!\!\!\!\!d\tau_1
\langle T_\tau\hat{d}_y(q,\tau)\hat{d}_y(q_1,\tau_1)\hat{d}_y(-q_1,\tau_1)\hat{d}_y(-q,0)\rangle_0\hskip1.5cm\nonumber\\
+\frac{1}{2}\Big(\frac{2\pi}{\hbar v_0}\Big)^{\!2}\!\sum_{q_1,q_2}\int_0^{\hbar\beta}\!\!\!\!\!d\tau_1\int_0^{\hbar\beta}\!\!\!\!\!d\tau_2
\langle T_\tau\hat{d}_y(q,\tau)\hat{d}_y(q_1,\tau_1)\hat{d}_y(-q_1,\tau_1)\hat{d}_y(q_2,\tau_2)\hat{d}_y(-q_2,\tau_2)\hat{d}_y(-q,0)\rangle_0-\cdots\nonumber.
\end{eqnarray}
This, after the pairing of the induced dipole operators per Wick's theorem with \emph{only} the connected Feynman diagram terms retained, gives
\begin{eqnarray}
\langle T_\tau S(\beta)\hat{d}_y(q,\tau)\hat{d}_y(-q,0)\rangle_0=\langle T_\tau\hat{d}_y(q,\tau)\hat{d}_y(-q,0)\rangle_0
-\!\!\int_0^{\hbar\beta}\!\!\!\!\!d\tau_1\langle T_\tau\hat{d}_y(q,\tau)\hat{d}_y(-q,\tau_1)\rangle_0\frac{2\pi}{\hbar v_0}\,\langle T_\tau\hat{d}_y(q,\tau_1)\hat{d}_y(-q,0)\rangle_0\nonumber\\
+\!\int_0^{\hbar\beta}\!\!\!\!\!d\tau_1\!\!\int_0^{\hbar\beta}\!\!\!\!\!d\tau_2\langle T_\tau\hat{d}_y(q,\tau)\hat{d}_y(-q,\tau_1)\rangle_0\frac{2\pi}{\hbar v_0}\,
\langle T_\tau\hat{d}_y(q,\tau_1)\hat{d}_y(-q,\tau_2)\rangle_0\frac{2\pi}{\hbar v_0}\,\langle T_\tau\hat{d}_y(q,\tau_2)\hat{d}_y(-q,0)\rangle_0-\cdots.\hskip1cm
\label{PyyexpansionWick}
\end{eqnarray}
In here, it is reflected that \emph{only} the summands with $q_1\!=q_2\!=\cdots=q$ produced by the pairing as shown can contribute nonzero trace values for the trace taken over the complete set of states (\ref{HilbertSpace}) of the noninteracting exciton Hilbert space. Substituting Eq.~(\ref{PyyexpansionWick}) in Eq.~(\ref{Pyynew}) and using the Matsubara (complex-time) function Fourier transform properties for bosons (see Ref.~\cite{Mahan2000}), summarized in our case as follows
\begin{eqnarray}
P_{yy}(q,i\omega)=\int_{0}^{\hbar\beta}\!\!\!\!\!d\tau\,e^{i\omega\tau}P_{yy}^{(0)}(q,\tau),\;\;\;\omega=\omega_n=\frac{2\pi n}{\hbar\beta},\;n=0,\pm1,\pm2,\cdots,\nonumber\\[0.25cm]
P_{yy}^{(0)}(q,\tau-\tau^\prime)=\frac{1}{\hbar\beta}\sum_{\omega_n}e^{-i\omega_n(\tau-\tau^\prime)}P_{yy}^{(0)}(q,i\omega_n),\hskip1.75cm\nonumber\\
\frac{1}{\hbar\beta}\int_{0}^{\hbar\beta}\!\!\!\!\!d\tau\,e^{i(\omega_n\!-\omega_{n^\prime})\tau}=\delta_{\omega_n\omega_{n^\prime}},\;\;\;
\frac{1}{\hbar\beta}\sum_{\omega_n}e^{-i\omega_n(\tau-\tau^\prime)}=\delta(\tau-\tau^\prime),\hskip0.75cm\nonumber
\end{eqnarray}
one obtains an infinite series
\begin{eqnarray}
P_{yy}(q,i\omega)=P_{yy}^{(0)}(q,i\omega)+P_{yy}^{(0)}(q,i\omega)\frac{2\pi}{v_0}P_{yy}^{(0)}(q,i\omega)+P_{yy}^{(0)}(q,i\omega)\frac{2\pi}{v_0}P_{yy}^{(0)}(q,i\omega)\frac{2\pi}{v_0}P_{yy}^{(0)}(q,i\omega)+\cdots,\nonumber
\end{eqnarray}
which can be rewritten in the closed form
\[
P_{yy}(q,i\omega)=P_{yy}^{(0)}(q,i\omega)+P_{yy}^{(0)}(q,i\omega)\frac{2\pi}{v_0}P_{yy}(q,i\omega)
\]
that results in Eq.~(\ref{Pyy}).

\twocolumngrid

\end{document}